\documentclass[submit]{aiaa-tc}
\usepackage[utf8]{inputenc}
\usepackage[hidelinks]{hyperref}

\usepackage{amsmath}
\usepackage{amssymb}
\usepackage{stmaryrd}
\usepackage{array}
\usepackage{xfrac}

\NewDocumentCommand \dx { O{x} } {\,\mathrm{d} #1}
\NewDocumentCommand \vect { m } { \mathbf{#1} }
\NewDocumentCommand \jump { m } { \left\llbracket #1 \right\rrbracket }

\title{A Comprehensive Study of Adjoint-Based Optimization of Non-Linear
Systems with Application to Burgers' Equation}

\author{
Alexandru Fikl, Vincent Le Chenadec, Taraneh Sayadi \\
{\normalsize\itshape
University of Illinois at Urbana-Champaign, Aerospace Engineering Department, United States}\\
\and
Peter J. Schmid\\
{\normalsize\itshape
Department of Mathematics, Imperial College London, United Kingdom}
}

\AIAApapernumber{2016}
\AIAAconference{AIAA AVIATION 2016, 13--17 June 2016, Washington, D.C.}
\AIAAcopyright{\AIAAcopyrightD{2016}}

\begin{document}

\maketitle

\begin{abstract}
In the context of adjoint-based optimization, nonlinear conservation laws
pose significant problems regarding the existence and uniqueness of both direct
and adjoint solutions, as well as the well-posedness of the problem for
sensitivity analysis and gradient-based optimization algorithms. In this paper
we will analyze the convergence of the adjoint equations to known exact
solutions of the inviscid Burgers' equation for a variety of
numerical schemes. The effect of the non-differentiability of the underlying
approximate Riemann solver, complete vs. incomplete differentiation of the
discrete schemes and inconsistencies in time advancement will be discussed.
\end{abstract}

\section{Introduction}

Adjoint methods have been traditionally used in flow and design optimization
(with seminal papers by Pironneau~\cite{P1974} and Jameson~\cite{J1988})
and sensitivity analysis, generally for linear equations, continuous flow
variables or numerical models that regularize shocks in some way (steady-state
RANS models, artificial viscosity, etc.). In the field of fluid
mechanics, adjoint methods are of particular interest because they
drastically reduce the size of the problem by using the \emph{adjoint
equations} instead of differentiation techniques to compute the derivatives
of a given cost functional (see~\cite{G2003} for a general introduction to flow
control). Optimization problems with PDE-based constraints have been the
focus of significant research, see, for example,~\cite{BS2012, HPUU2008} for a
study with an emphasis on elliptic/parabolic equations, and~\cite{L1971}
for a view on hyperbolic equations as optimization constraints. Less often
studied, but still of great importance, are applications that require the use
of nonlinear PDEs with highly localized features (reactive zones), large
gradients (boundary layers), discontinuities (shocks) or interfaces (multiphase
flows). These complexities,
which could even arise for well-behaved initial data, pose significant problems
to the existence and uniqueness of a solution to the state and the adjoint
equations, which can have tremendously negative effects on the convergence of
any optimization algorithm.

The issues of adjoint optimization in the context of nonlinear conservation
laws with shocks have been studied numerically to a large extent in the
past (see~\cite{ZZANS2000, IS1999, MH1997}). Non-differentiable optimization
techniques, based on subgradients, have been successfully employed to solve
such problems~\cite{HN2003}. There have also been attempts at using
methods which do not rely on the gradient information, such as stochastic
optimization~\cite{HL2001} or genetic algorithms~\cite{OONN1997}. However, these
methods also suffer from limitations, usually related to very high computational
costs compared to classical gradient descent methods.

Despite the importance of problems involving some kind of discontinuity
in the flow variables, little attention has been placed on the existence
and uniqueness or the consistency and convergence of the continuous or discrete
equations, respectively. Recent contributions such as~\cite{BP2003, G2002}
and~\cite{GU12010, GU22010} have formally defined
the problems and given proofs of
the existence of an optimal control, the pointwise convergence almost everywhere
of the linearized and adjoint equations, as well as of the cost functional in
the case of scalar conservation laws. These
proofs have been extended to cases with Dirac initial data and multiple shocks
and provide a solid basis for future analysis. Another approach that has
received attention is the possibility of adding the shock locations as
independent variables to the existing set of equations~\cite{CHS1997}, and
developing specialized descent algorithms that use this information~\cite{CPZ2008}.
Entirely new numerical methods that explicitly track
the shock location may be required as well, which could be too complex to be
practical in some cases (see also the work on front-tracking in~\cite{GIMM1981}).

In this paper, we propose to perform extensive numerical studies on various
discrete schemes for nonlinear conservation laws and provide insights into
their convergence (if possible) in different scenarios. This follows the numerical
work done in~\cite{GU12010, GU22010} on a linear modified Lax-Friedrichs
scheme, where convergence has been formally proven for both the linearized
equations and the adjoint equations. However, the analytical work done by
Giles and Ulbrich is extremely difficult to extend to more advanced nonlinear
numerical schemes, given the current corpus of knowledge in the field of
stability and convergence analysis. Thus, we propose rigorous numerical testing
to gain insight into the behavior of widely used nonlinear numerical schemes.

There are many issues that can be investigated in this area. Firstly, we will
investigate the time discretization of the adjoint equations, as obtained by
taking the transpose of the linearized discrete operators. When
discretizing the continuous adjoint equations, it is unclear at what time
the state variables that appear in nonlinear schemes should be taken, e.g.
$t^m, t^{m + 1}$ or some intermediate level in multi-stage time integration.
This issue will be discussed in Section~\ref{sec:time} in the case of discrete
adjoint equations, which should shed light on the continuous case as well.

We will also look, in Section~\ref{sec:space}, at the spatial discretization and
see what impact non-differentiable approximate Riemann solvers will have
on the convergence of the numerical schemes. This issue is mostly of relevance
when automatic differentiation software is not used and the discrete equations
are differentiated manually. It has been suggested in~\cite{AP2012} that
incomplete differentiation has very little impact on the end results (for the
Euler equations with a Roe flux).

Following these investigations, in Section~\ref{sec:schemes}, we will use the
modified Lax-Friedrichs scheme proposed in~\cite{GU12010, GU22010} to construct
a hybrid numerical scheme that uses a second order flux limited scheme away from
the shocks, resulting in the expected convergence at the shock and high-order
accuracy in smooth regions. A major finding by~\cite{GU12010, G2002}
concerning converging schemes, like the modified Lax-Friedrichs scheme, is that
they generally require an inherent form of artificial dissipation.
The dissipation has the effect of smearing the shock across an
increasing number of cells as the grid is refined, thus allowing the numerical
scheme to implicitly satisfy the boundary conditions for the adjoint at the
shock locations. This remark may also explain why many practical tests that
have been done in the past, using artificial viscosity or similar techniques,
have been successful to quite a large extent.

\section{Adjoint Methods} \label{sec:adjoint}

There are two main branches of adjoint methods in use today: the
\emph{differentiate-then-discretize} approach, which derives the continuous
adjoint equations from the state equations and discretizes them separately,
and the \emph{discretize-then-differentiate} approach, which discretizes the state
equations and derives a discrete adjoint that is consistent with the discrete
state equations. The first approach is usually preferred because it allows
different choices of numerical schemes or even meshes for the state
and adjoint equations. This can prove very advantageous, for example, when
using adaptively refined meshes, since the state and adjoint variables will
have very different areas of interest. A downside to the
\emph{differentiate-then-discretize} approach is that, at a discrete level,
the obtained gradient is not guaranteed to be consistent, and typically gives rise
to numerical instabilities. On the other hand, in the
\emph{discretize-then-differentiate} approach, the adjoint equations are
obtained from the discrete state equations, and the gradient is guaranteed to
be consistent at the discrete level. For a more general discussion on the
trade-offs involved see~\cite{G2003, NJ2000}.

\subsubsection*{Lagrangian Formalism}

We will first look at the continuous case and the
\emph{differentiate-then-discretize} approach. A common formulation for
constrained optimization problems is as follows:
\begin{equation} \label{eq:minimization}
\left\{
\begin{aligned}
& \min_{u, g} \mathcal{J}(u, g), \\
& F(u, g) = 0, \\
& g \in \mathcal{G}_\mathrm{ad},
\end{aligned}
\right.
\end{equation}
where $u$ is known as a \emph{state variable}, $g$ is the \emph{control variable},
$F(u, g)$ is a set of (PDE-based) constraints and $\mathcal{G}_{\mathrm{ad}}$
is a set of admissible controls. Assuming that the functional $\mathcal{J}$
allows a unique minimizer $g^*$ (see, for example,~\cite{BS2012} for very general
considerations), one way to solve such a \emph{constrained optimization problem}
is to transform it into an unconstrained optimization problem using
\emph{Lagrange relaxation} and the Lagrangian functional:
\[
\mathcal{L}(u, g, p) = \mathcal{J}(u, g) - <p, F(u, g)>,
\]
where $p$ is an \emph{adjoint} variable (usually referred to as a Lagrange
multiplier) introduced to enforce the constraints $F(u, g)$ and
$<\cdot, \cdot>$ is an appropriate inner product. If we assume that our
variables $(u, g)$ are smooth, we can define the well-known first order
necessary conditions for the minimizer of $\mathcal{L}$:
\begin{subequations}
\begin{align}
& 0 = \frac{\partial \mathcal{L}}{\partial p}(u, g, p) = -F(u, g),
    \label{eq:optimality_state} \\
& 0 = \frac{\partial \mathcal{L}}{\partial u}(u, g, p) =
    \frac{\partial \mathcal{J}}{\partial u} -
    \left(\frac{\partial F}{\partial u}\right)^* p,
    \label{eq:optimality_adjoint} \\
& 0 = \frac{\partial \mathcal{L}}{\partial g}(u, g, p) =
    \frac{\partial \mathcal{J}}{\partial g} -
    \left(\frac{\partial F}{\partial g}\right)^* p,
    \label{eq:optimality_cond}
\end{align}
\end{subequations}
which form the \emph{optimality system}, where~\eqref{eq:optimality_state}
retrieves the constraints,~\eqref{eq:optimality_adjoint} is the adjoint system
and~\eqref{eq:optimality_cond} is the so-called optimality condition. We have
denoted by $(\cdot)^*$ the adjoint of the given operator. However, the ultimate
goal is to find a formula for the derivative of $\mathcal{J}(u(g), g)$ with
respect to $g$ and, thus, the sensitivity of the cost functional with respect to
each control variable. The chain rule gives:
\begin{equation} \label{eq:cost_derivative_orig}
\frac{\mathrm{d} \mathcal{J}}{\mathrm{d} g} =
\frac{\partial \mathcal{J}}{\partial g} +
\frac{\partial \mathcal{J}}{\partial u} \frac{\mathrm{d} u}{\mathrm{d} g}.
\end{equation}
If we differentiate the constraints $F(u(g), g)$, we find:
\begin{equation} \label{eq:constraint_derivative}
\frac{\mathrm{d} F}{\mathrm{d} g} =
\frac{\partial F}{\partial u} \frac{\mathrm{d} u}{\mathrm{d} g} +
\frac{\partial F}{\partial g} = 0
\implies
\frac{\mathrm{d} u}{\mathrm{d} g} =
\left(\frac{\partial F}{\partial u}\right)^{-1} \frac{\partial F}{\partial g}.
\end{equation}

We can then substitute~\eqref{eq:optimality_adjoint}
and~\eqref{eq:constraint_derivative} into~\eqref{eq:cost_derivative_orig} to
get a formula for the derivative:
\begin{equation} \label{eq:cost_derivative}
\frac{\mathrm{d} \mathcal{J}}{\mathrm{d} g} =
\frac{\partial \mathcal{J}}{\partial g} -
\left(\frac{\partial F}{\partial g}\right)^* p.
\end{equation}

Equation~\eqref{eq:cost_derivative} illustrates the great advantage of using
adjoint methods in optimization. In~\eqref{eq:cost_derivative_orig}, we would
need to compute the sensitivity of the state variables with respect to each
control, which would lead to solving $K$ systems of sensitivity equations (where
$K$ is the number of control variables, assuming~\eqref{eq:minimization} describes
a finite-dimensional system, upon discretization of the original PDE).
However, in~\eqref{eq:cost_derivative},
we only need to solve the adjoint equation once and then we can get all
the components of the derivative of the cost function $\mathcal{J}$.

It should be noted that~\eqref{eq:cost_derivative} is written in terms of directional
derivatives. To retrieve the gradient, an appropriate inner product has to
be chosen, resulting from a straightforward application of the Riesz
representation theorem (see~\cite{G2003, BS2012} for the implications stemming
from the choice of inner product).

\subsubsection*{Linear Algebra Formalism}

In the case of the \emph{differentiate-then-discretize} approach to adjoint
methods, it is perhaps more intuitive to look at the problem from a linear
algebra point of view, since the variables $(\vect{u}, \vect{g})$ are now
discrete values approximated at each grid point and time step. We also denote by
$\mathcal{J}^h(\vect{u}, \vect{g})$ the discretized cost function and by
$F^h(\vect{u}, \vect{g})$ the discretized constraints, using any numerical
scheme, in both space and time. We can look at a discrete  equivalent of the
derivative of the cost function from~\eqref{eq:cost_derivative_orig}:
\[
\frac{\mathrm{d} \mathcal{J}^h}{\mathrm{d} \vect{g}} =
\frac{\partial \mathcal{J}^h}{\partial \vect{g}} +
\frac{\partial \mathcal{J}^h}{\partial \vect{u}}
\frac{\mathrm{d} \vect{u}}{\mathrm{d} \vect{g}}
\]
and the discretized constraints from~\eqref{eq:constraint_derivative}:
\[
\frac{\mathrm{d} F^h}{\mathrm{d} \vect{g}} =
\frac{\partial F^h}{\partial \vect{u}}
\frac{\mathrm{d} \vect{u}}{\mathrm{d} \vect{g}} +
\frac{\partial F^h}{\partial \vect{g}}.
\]

The two equations can be formulated as:
\begin{equation} \label{eq:adjoint_algebra}
\frac{\mathrm{d} \mathcal{J}^h}{\mathrm{d} \vect{g}} =
\frac{\partial \mathcal{J}^h}{\partial \vect{g}} +
\vect{h}^T \vect{v}
\quad \text{with} \quad
L \vect{v} = \vect{f},
\end{equation}
where we have renamed the variables as follows:
\[
\begin{aligned}
\vect{v} =~ & \frac{\mathrm{d} \vect{u}}{\mathrm{d} \vect{g}},
& \quad \quad
\vect{h}^T =~ & \frac{\partial \mathcal{J}^h}{\partial \vect{u}}, \\
L =~ & \frac{\partial F^h}{\partial \vect{u}},
& \quad \quad
\vect{f} =~& -\frac{\partial F^h}{\partial \vect{g}}.
\end{aligned}
\]

From a linear algebra point of view, the above variables are simple matrices
and vectors and we can simply introduce a new variable $\vect{p}$ that will
allow us to replace $\vect{h}^T \vect{v}$ with $\vect{p}^T \vect{f}$, where
$\vect{p}$ must satisfy $L^T \vect{p} = \vect{h}$. The fact that the two
terms are indeed equivalent can be easily seen from:
\[
\vect{p}^T \vect{f} =
\vect{p}^T L \vect{v} = (L^T \vect{p})^T \vect{v} =
\vect{h}^T \vect{v}.
\]

A more thorough overview of the linear algebra viewpoint is given in~\cite{GP2000}.
This formalism is particularly appealing in the case of sensitivity and
error analysis, where invoking the Lagrangian formalism, widely used in
optimization, may seem too complex. The two methods are, of course, completely
equivalent, but each has its strengths in particular contexts. As we will see,
the Lagrangian viewpoint may be more intuitive when analyzing discrete numerical
schemes (as it does not hide the complexity behind operators such as $L$),
while the linear algebra viewpoint can be used in defining a modular,
operator-based framework for the implementation of adjoint optimization
methods (see~\cite{FSS2012}).

\subsubsection*{Burgers' Equation and Shocks}

The model problem we will consider is the inviscid Burgers' equation:
\begin{equation} \label{eq:burgers}
\begin{cases}
\partial_t u + \partial_x f(u) = 0, & \quad (x, t) \in [a, b] \times [0, T], \\
u(x, 0) = g(x),                     & \quad x \in [a, b], \\
\end{cases}
\end{equation}
where $f(u) = u^2 / 2$ and the equation is supplemented by appropriate boundary
conditions on inflowing boundaries. For simplicity, we take the initial condition
$g$ as a control variable and a tracking-type cost functional:
\begin{equation} \label{eq:cost}
\mathcal{J}(u, g) = \int_a^b G(u(T)) \dx[x].
\end{equation}

Using~\eqref{eq:optimality_adjoint}, we derive the adjoint equations:
\begin{equation} \label{eq:adjoint}
\begin{cases}
-\partial_t p - u \partial_x p = 0, & \quad (x, t) \in [a, b] \times [0, T], \\
p(x, T) = G'(u(T)),                 & \quad x \in [a, b],
\end{cases}
\end{equation}
where $p$ is the adjoint variable, as before. Given the fact that the
Lagrangian is linear in $p$, the adjoint equations will themselves
be linear PDEs. However, this formal derivation of the adjoint is done under
strong continuity and differentiability hypotheses for all functionals and
variables. Given the nonlinear nature of~\eqref{eq:burgers}, shocks will
develop for smooth initial datum at the break time:
\[
T_b = \min_x \left(-\frac{1}{\partial_x u(x, 0)}\right),
\]
so we can no longer rely on the previous results for $T > T_b$ (provided
$T_b > 0$, of course). This case has been discussed in detail in~\cite{G2002}
and more recently in~\cite{GU12010, GU22010} and involves keeping track of
the shock position as defined by the Rankine-Hugoniot conditions across the shock:
\[
\dot{x}_s \jump{u} = \jump{f(u)},
\]
where $\dot{x}_s$ is the shock velocity (thus $x_s$ is its position) and
$\llbracket \cdot \rrbracket$ denotes the jump in the given quantity across
the shock. If we assume that the shock is present on a curve
$\Gamma = (t, x_s(t))$, then the equations we have defined above remain valid
for $\Omega \setminus \Gamma$ and a new set of equations has to be added to
define the perturbation to the shock position given by the Rankine-Hugoniot
condition. From~\cite{GU12010}, it is:
\begin{equation} \label{eq:shock}
\dot{\tilde{x}}_s \jump{u} +
\dot{x}_s \jump{\tilde{u} + \tilde{x}_s \frac{\partial u}{\partial x}} -
\jump{u\tilde{u} + \tilde{x}_s \frac{\partial f(u)}{\partial x}} = 0,
\end{equation}
which can be further reduced to the following evolution equation for the
linearized shock position:
\[
\frac{\mathrm{d}}{\mathrm{d}t} \left(\tilde{x}_s \jump{u}\right) =
\jump{(u - \dot{x}_s) \tilde{u}},
\]
where $\tilde{u}$ and $\tilde{x}_s$ are small perturbations to the base state
$(u, x_s)$.

The derivation above is largely based on the theory of generalized tangent vectors
developed in~\cite{BM1995} as a new form of variational calculus for such
systems. This methodology is also described in~\cite{GU12010, G2002} and
in~\cite{CPZ2008}, which proposes a new steepest descent method including
the shock positions in the system. A very important
point to be made is that, even if the state variables are discontinuous along
the shock, the adjoint variable $p$ in~\eqref{eq:adjoint} has a uniform value
along all characteristics leading backwards from the shock at position
$x_s$~\cite{G2002}.

A different formal approach is taken in~\cite{U2001}, where the author has
tried to contain the amount of knowledge needed about the shock positions
in formal convergence proofs. To this end, a new notion of
\emph{shift-differentiability} is defined for functions with moving
discontinuities, which implies, under fairly general assumptions, the
Fréchet-differentiability of tracking-type functionals as the one given
by~\eqref{eq:cost}. Furthermore,~\cite{U2001} shows that entropy solutions
to conservation laws, such as~\eqref{eq:burgers}, are shift-differentiable
under fairly general hypotheses, paving the way for a rigorous definition of
the adjoint equations in the presence of shocks.

We also note that, for Burgers' equation, the adjoint equation is an
advection equation with a discontinuous velocity. This case has been studied
in~\cite{BJ1998} and~\cite{GJ2000}, where it has been proven that there exist
unique solutions, called \emph{reversible solutions} for this type of equation
in the space of local Borel measures $\mathcal{M}_{\mathrm{loc}}(\mathbb{R})$.
Extensions to the case of adjoint equations were proposed in~\cite{U2001} which
can handle discontinuous end data, non-homogeneous right-hand sides, rarefactions
waves and, most importantly, prove the differentiability for the cost functional.

\section{Temporal Consistency} \label{sec:time}

The interest of this paper lies in studying the convergence of discrete
adjoint numerical schemes, i.e. schemes obtained by using the
\emph{discretize-then-differentiate} approach to adjoint optimization
(see~\cite{G2003}). To give a proof of concept for this approach, we first
discretize~\eqref{eq:burgers} using a classic Finite Volume method. The domain
$\Omega = [a, b]$ is discretized into $N$ cells:
\[
T_i = [x_{i - \frac{1}{2}}, x_{i + \frac{1}{2}}]
\]
centered at $x_i$ with a uniform cell size $\Delta x$. By integrating over each
space-time cell $T_i \times [t^m, t^{m + 1}]$, we get the following update
formula for the state variable $u$:
\begin{equation} \label{eq:burgers_fv}
\begin{cases} \displaystyle
u^{m + 1}_i = u^m_{i} - \frac{\Delta t}{\Delta x}
\left(f^m_{i + \frac{1}{2}} - f^m_{i - \frac{1}{2}}\right), \\
u^0_i = g_i,
\end{cases}
\end{equation}
where the cell and face averaged variables are defined as:
\[
\begin{aligned}
u^m_i & = \frac{1}{\Delta x} \int_{T_i} u(x, t^m) \dx[x], \\
f^m_{i - 1/2} & =
    \frac{1}{\Delta t} \int_{t^m}^{t^{m + 1}} f(u(x_{i - 1/2}, t)) \dx[t].
\end{aligned}
\]

We will now look at the modified Lax-Friedrichs scheme proposed in~\cite{GU12010},
for which the flux may be written as:
\begin{equation} \label{eq:lax_flux}
f^m_{i - \frac{1}{2}} = \frac{1}{2}(f(u^m_i) + f(u^m_{i - 1})) -
                        \frac{\epsilon}{\Delta x} (u^m_i - u^m_{i - 1}),
\end{equation}
where $\epsilon = \Delta x^\alpha$ and $\alpha \in (\sfrac{2}{3}, 1)$ is a
grid-dependent parameter that will smear the shock over an increasing number of
cells as $\Delta x \to 0$. The proposed scheme is stable under the CFL condition:
\[
\max_i |u^m_i| \epsilon \frac{\Delta t}{\Delta x^2} \le \frac{1}{2}.
\]

Using these definitions, the update formula given in~\eqref{eq:burgers_fv}
describes a differentiable (with respect to $u^m_i$) finite volume scheme
for~\eqref{eq:burgers} of order $\mathcal{O}(\epsilon)$. It is important to
note that, in the continuous case, a formal analysis requires keeping
track of the shock positions with additional equations like~\eqref{eq:shock},
but in the discrete case, the complexity can be sidestepped if the numerical
scheme smears the shock appropriately~\cite{G2002}.

If we denote by $\vect{u} = \left ( u^m_i \right )$ and $\vect{g} = \left ( g_i \right )$
\[
g_i = \frac{1}{\Delta x} \int_{T_i} g(x) \dx[x],
\]
the solution vector and the control variable, respectively.
The discretized cost functional~\eqref{eq:cost} can be written as:
\[
\mathcal{J}^h(\vect{u}, \vect{g}) = \sum_i \Delta x\, G(u^M_i).
\]

Following the results from Section~\ref{sec:adjoint}, the discrete adjoint
equations are:
\begin{equation} \label{eq:adjoint_fv}
\begin{cases} \displaystyle
p^{m}_i = p^{m + 1}_i +
\frac{\Delta t}{\Delta x}
\left(f^{*, m + 1}_{i, +} + f^{*, m + 1}_{i, -}\right), \\
p^{M}_i = G'(u^M_i),
\end{cases}
\end{equation}
where $M$ denotes the total number of time steps and the ``fluxes'' are given by:
\begin{equation} \label{eq:adjoint_flux}
\left\{
\begin{aligned}
f^{*, m + 1}_{i, -} =~ &
\frac{1}{2} u^m_i (p^{m + 1}_i - p^{m + 1}_{i - 1}) -
\frac{\epsilon}{\Delta x} (p^{m + 1}_i - p^{m + 1}_{i - 1}), \\
f^{*, m + 1}_{i, +} =~ &
\frac{1}{2} u^m_i (p^{m + 1}_{i + 1} - p^{m + 1}_i) +
\frac{\epsilon}{\Delta x} (p^{m + 1}_{i + 1} - p^{m + 1}_i). \\
\end{aligned}
\right.
\end{equation}

We have written the discrete system in a very specific way so that it
invites comparison to the continuous case~\eqref{eq:adjoint}. However,
discretely, the adjoint equation is simply given by the
transpose of the linearized discrete operator. In order to clearly see the
fluxes, we consider the equations on a per-time step basis instead:
\begin{equation} \label{eq:discrete_space_operator}
\vect{u}^{m + 1} = \vect{u}^m - \mathcal{A}(\vect{u}^m),
\end{equation}
where $\mathcal{A}$ contains the flux term in~\eqref{eq:burgers_fv}. The
adjoint equation is then simply obtain by linearizing $\mathcal{A}$ around
a reference state $\vect{u}^m$ and transposing the operator to obtain:
\[
\vect{p}^{m} = \vect{p}^{m + 1} -
\left(\frac{\partial \mathcal{A}}{\partial \vect{u}^m}\right)^T \vect{p}^{m + 1}.
\]

Using this formulation it is perhaps more intuitive to write the adjoint
equation as follows:
\begin{equation} \label{eq:adjoint_fv_general}
p^m_i = p^{m + 1}_i -
\frac{\Delta t}{\Delta x}
\left[
\left(\frac{\partial f^m_{i - \sfrac{1}{2}}}{\partial u^m_i}\right) p^{m + 1}_{i - 1}
+
\left(
\frac{\partial f^m_{i + \sfrac{1}{2}}}{\partial u^m_i} -
\frac{\partial f^m_{i - \sfrac{1}{2}}}{\partial u^m_i}
\right) p^{m + 1}_i
+
\left(-\frac{\partial f^m_{i + \sfrac{1}{2}}}{\partial u^m_i}\right) p^{m + 1}_{i + 1}
\right],
\end{equation}
where it is obvious that the second term on the right-hand side is the result
of a matrix-vector multiplication. In this very specific case, the operator
$\mathcal{A}$ only has 3 elements on row $i$ at positions $(i - 1, i, i + 1)$
corresponding  to the stencil of the numerical scheme, so we see only 3 terms
in~\eqref{eq:adjoint_fv_general}. This is a very generic formulation, as opposed
to~\eqref{eq:adjoint_flux}, where the fluxes have already been differentiated,
with obvious extensions to schemes with wider stencils and more complicated
flux formulæ.

It should be noted that, in the discrete case, no specific attention needs to be
set on the differentiability and the presence of shocks since we are
differentiating in $\mathbb{R}^n$. However, we do need to worry about the
differentiability of the numerical scheme and, more importantly, the convergence
to the continuous equations, which is itself a very complex issue. Numerical
and formal proofs of convergence have been given in~\cite{GU12010} for the
modified Lax-Friedrichs scheme adopted here.

We will now briefly look at the convergence and the order of accuracy for the case
of the modified Lax-Friedrichs scheme~\eqref{eq:burgers_fv} and its
adjoint~\eqref{eq:adjoint_fv}. For testing purposes, we take a cost functional
with $G(u) = \frac{1}{2} u^2$ and the following exact solution with a single
shock at $x = 0$:
\begin{equation} \label{eq:burgers_exact_1}
u(x, t) =
\begin{cases}
1.5,    & \quad x < 0.5 t,\\
-0.5,   & \quad x > 0.5 t
\end{cases}
\end{equation}
and a known solution to the adjoint equation~\eqref{eq:adjoint} for $T = 1$:
\begin{equation} \label{eq:burgers_adjoint_exact_1}
p(x, t) =
\begin{cases}
1.5,    & \quad x < -1 + 1.5 t, \\
0.5,    & \quad -1 + 1.5 t < x < 1 - 0.5 t, \\
-0.5,   & \quad x > 1 + 0.5 t,
\end{cases}
\end{equation}
where the middle state in the adjoint is computed from $\jump{G(u)} / \jump{u}$
(see~\cite{G2002}). We also take $T = 1.0$, $\Omega = [-1.5, 1.5]$, and a CFL of
$0.9$. A similar test has been performed in~\cite{AP2012} for the upwind scheme
and very good convergence results were obtained. We can see in
Table~\ref{tbl:ulbrich_convergence} that the forward scheme converges with
the expected order $\alpha$, while the adjoint equation converges with order
$\sfrac{\alpha}{2}$. We have used the following discrete $L_1$ norm for the
error estimates:
\[
\|e_*\|_1 = \sum \Delta x\, |u^M_i - u(x, T)|,
\]
which aggregates both time and space errors.

\begin{table}[ht!]
\caption{$L_1$ error results for the state variable $e_u$ and the adjoint
variables $e_p$.}
\def\arraystretch{1.2}
\begin{center}
\begin{tabular}{lcccc}
  & 128 & 256 & 512 & \textbf{Order} \\\hline\hline
$\log(\|e_u\|_1)$ ($\alpha = 0.999$)
    & -2.77028 & -3.42418 & -4.15565 & \textbf{0.999} \\
$\log(\|e_p\|_1)$ ($\alpha = 0.999$)
    & -1.13771 & -1.49982 & -1.85148 & \textbf{0.514} \\\hline
$\log(\|e_u\|_1)$ ($\alpha = 0.9$)
    & -2.37230 & -2.98675 & -3.61683 & \textbf{0.897} \\
$\log(\|e_p\|_1)$ ($\alpha = 0.9$)
    & -0.91325 & -1.21177 & -1.52599 & \textbf{0.441} \\\hline
$\log(\|e_u\|_1)$ ($\alpha = 0.8$)
    & -1.99073 & -2.54112 & -3.09571 & \textbf{0.797} \\
$\log(\|e_p\|_1)$ ($\alpha = 0.8$)
    & -0.72734 & -0.96850 & -1.24296 & \textbf{0.372} \\\hline
\end{tabular}
\label{tbl:ulbrich_convergence}
\end{center}
\end{table}

\subsection{First Order Explicit Euler Time Integration}

We now investigate the time consistency of the adjoint equations.
Namely, in the adjoint fluxes~\eqref{eq:adjoint_flux}, we have used the state
variables at time $t^m$, while the adjoint variable was advancing backwards in
time from $t^{m + 1}$ to $t^m$. This is not obvious from the
continuous formulation and hidden beneath the linearized discrete operator $L$
in~\eqref{eq:adjoint_algebra}. To better understand the problem, we will look
at a scalar ODE of the form:
\[
y'(t) = f(t),
\]
which, discretized using an explicit forward Euler integration method, is:
\[
\frac{y^{m + 1} - y^m}{\Delta t} = f(t^m).
\]
We can also write the above equation as $A \vect{y} = \vect{f}$. In the
continuous case, the adjoint of $\partial_t$ is $-\partial_t$. However, in
the discrete case the time derivative operator $A$ and its adjoint $A^T$ are
given by:
\[
A = \frac{1}{\Delta t}
\begin{bmatrix}
1   &   0   &   0   & \cdots&   0   \\
-1  &   1   &   0   & \cdots&   0   \\
0   &  -1   &   1   & \cdots&   0   \\
\vdots&     &\ddots &       &\vdots \\
 0  &       &  -1   &   1   &   0   \\
 0  &\cdots &   0   &  -1   &   1
\end{bmatrix}
\quad \implies \quad
A^T = \frac{1}{\Delta t}
\begin{bmatrix}
1   &  -1   &   0   & \cdots&   0   \\
0   &   1   &   -1  & \cdots&   0   \\
0   &   0   &   1   & \cdots&   0   \\
\vdots&     &\ddots &       &\vdots \\
 0  &       &   0   &   1   &   -1  \\
 0  &\cdots &   0   &  0    &   1
\end{bmatrix}
\]

We will now look at the $m$-th term of the dot products $<\vect{x}, A\vect{y}>$
and $<A^T\vect{x}, \vect{y}>$, where $\vect{y}$ is our variable and $\vect{x}$
its adjoint. We have:
\[
\left\{
\begin{aligned}
x^m & \left(-\frac{1}{\Delta t} y^{m - 1} + \frac{1}{\Delta t} y^m\right), \\
y^m & \left(\frac{1}{\Delta t} x^m - \frac{1}{\Delta t} x^{m + 1}\right),
\end{aligned}
\right.
\quad \Longleftrightarrow \quad
\left\{
\begin{aligned}
x^m & \left(\frac{y^m - y^{m - 1}}{\Delta t}\right), \\
y^m & \left(-\frac{x^{m + 1} - x^m}{\Delta t}\right),
\end{aligned}
\right.
\]
which is also consistent with the property of the continuous operators. This
equality shows that the $m$th term from $<A^T \vect{x}, \vect{y}>$ matches
the ``state'' variable $y$ at time $t^m$ with the adjoint variable
$x$ advancing backwards in time from $t^{m + 1}$ to $t^m$, as expected.

\begin{figure}[ht!]
\centering
\begin{minipage}{0.5\textwidth}
\centering
\includegraphics[width=1.05\linewidth]{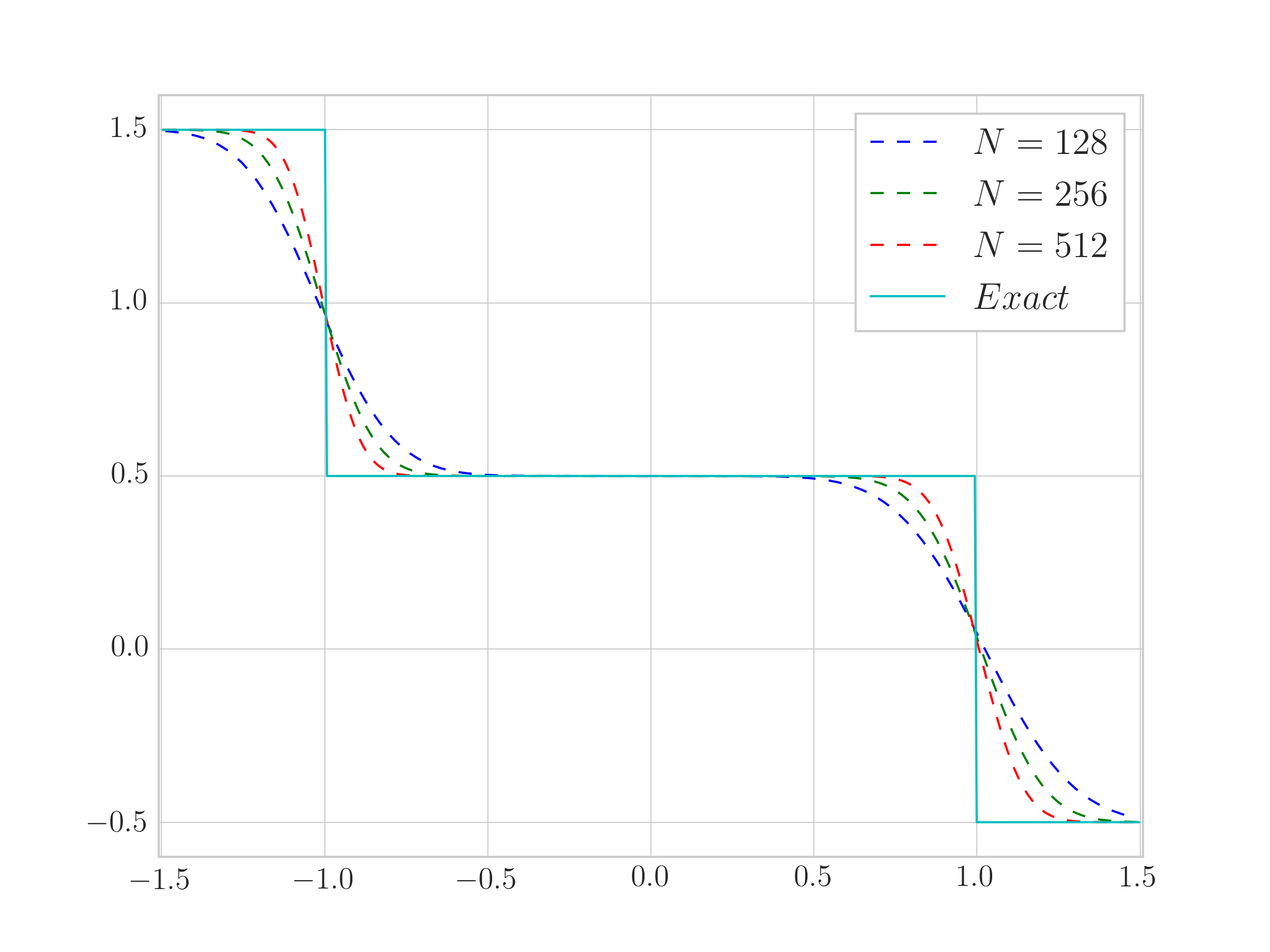}
\end{minipage}%
\begin{minipage}{0.5\textwidth}
\centering
\includegraphics[width=1.05\linewidth]{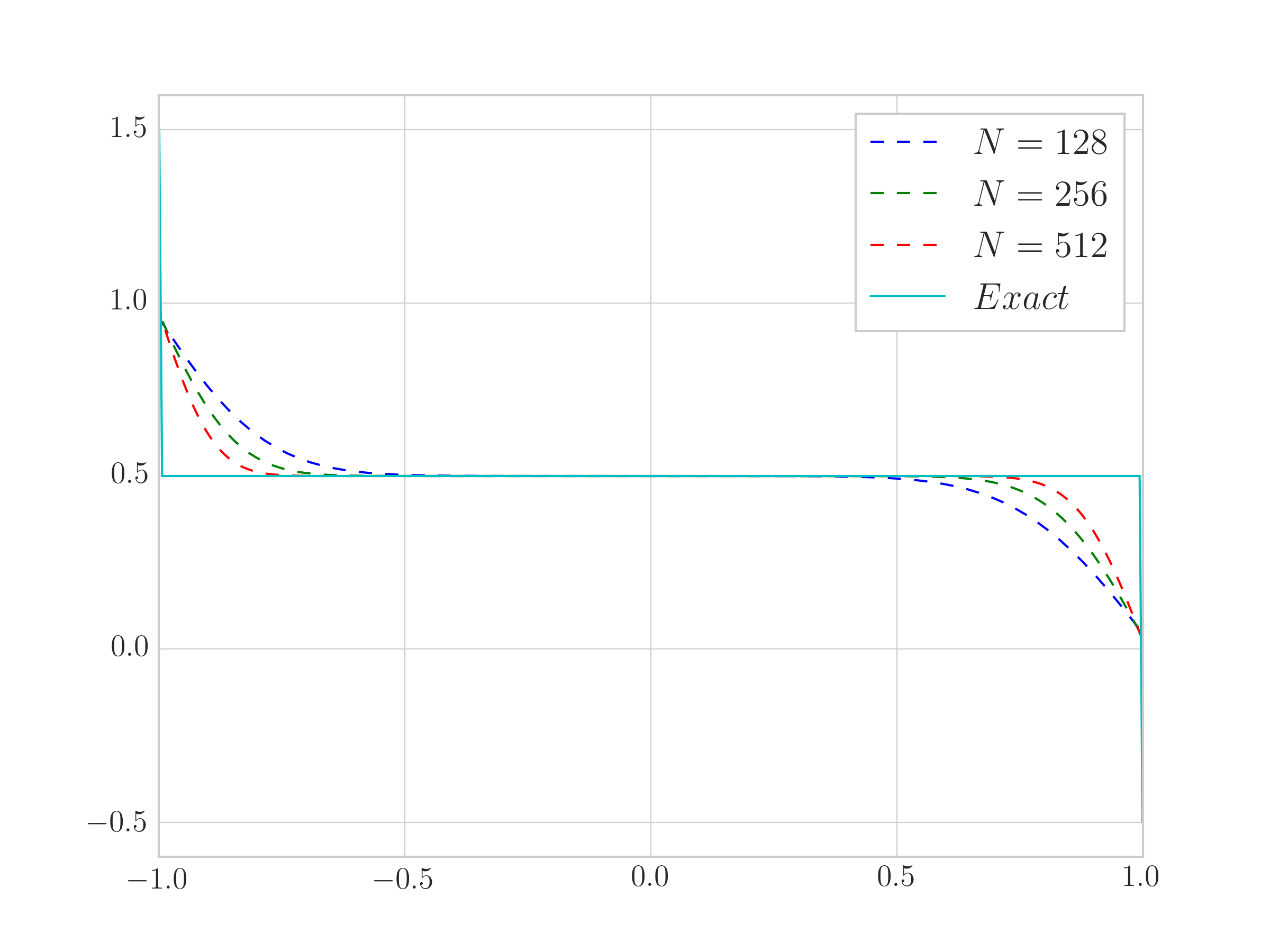}
\end{minipage}
\caption{Adjoint solution at $t = 0$ for~\eqref{eq:burgers_adjoint_exact_1}
(left) and a zoom around the shock position at $x = 0$ (right) with
consistent time-stepping.}
\label{fig:ulbrich_time_correct}
\end{figure}

\begin{figure}[ht!]
\centering
\includegraphics[width=0.55\linewidth]{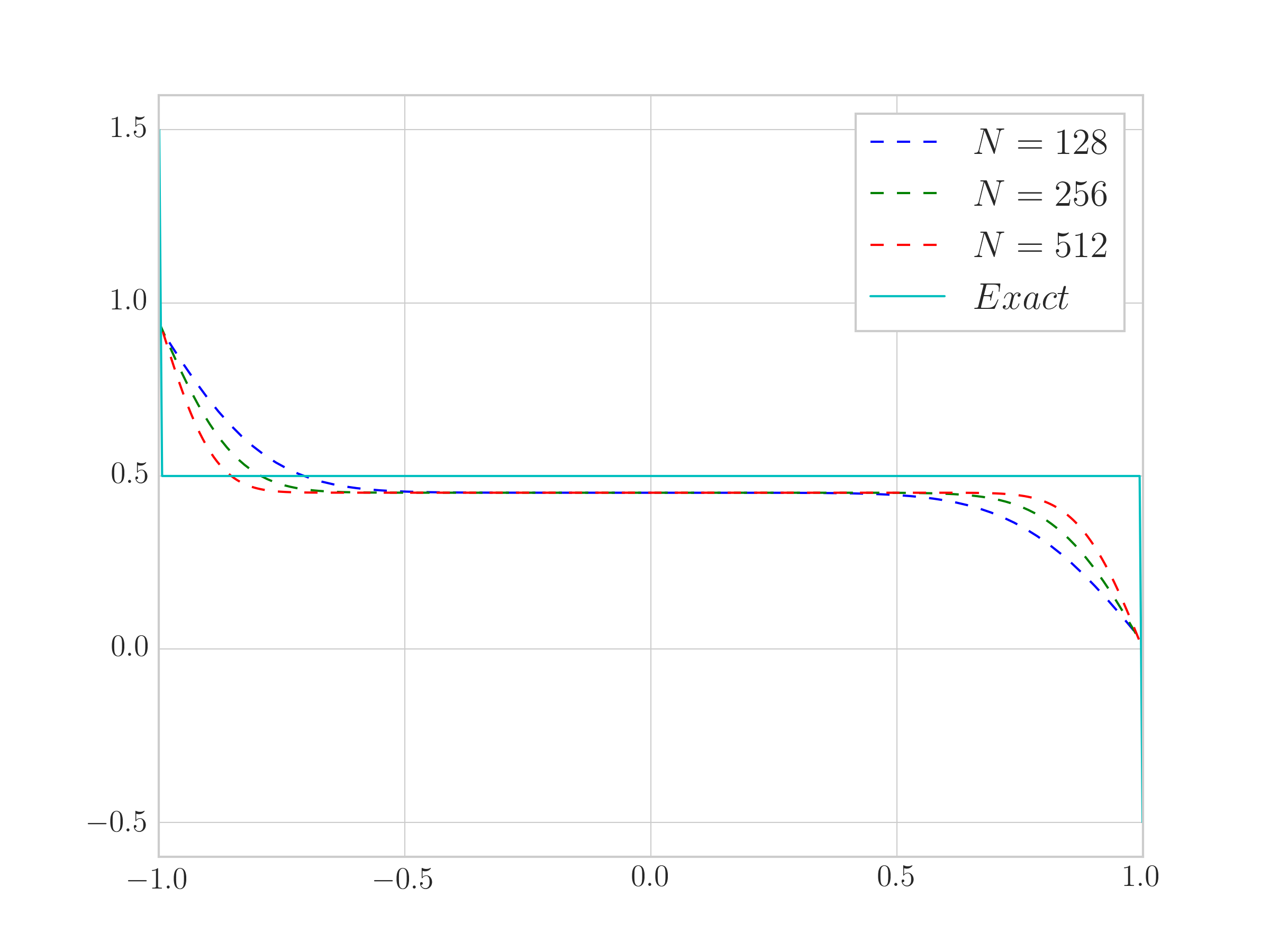}
\caption{Adjoint solution at $t = 0$ for~\eqref{eq:burgers_adjoint_exact_1}
zoomed around the shock position at $x = 0$.}
\label{fig:ulbrich_time_wrong}
\end{figure}

Figure~\ref{fig:ulbrich_time_correct} shows that the modified Lax-Friedrichs
scheme with the consistent time advancement converges correctly to the exact
adjoint solution. Note that this is not exactly the case for the upwind scheme
in~\cite{AP2012} where the solution seems to oscillate slightly around the
exact solution. Figure~\ref{fig:ulbrich_time_wrong} illustrates the
results of the scheme using $u^{m + 1}_i$ in~\eqref{eq:adjoint_flux}. The
scheme no longer converges to the exact solution and the plateau is slightly
below the exact value. This is due to the inconsistent time integration
scheme used.

\subsection{Second-Order Runge-Kutta Time Integration}

While the issues regarding time integration of the discrete adjoint are rather
obvious for the explicit Euler method used above, they are more subtle for
higher order methods, such as the Runge-Kutta family of methods. In this section
we will analyze two second-order Runge-Kutta methods and their adjoints, one
explicit and one implicit method. Generally, a time integrator $\phi_{\Delta t}$
has a classic definition of its adjoint~\cite{HWL2006}:
\begin{equation} \label{eq:rk_adjoint}
\phi^*_{\Delta t} = \phi^{-1}_{-\Delta t},
\end{equation}
i.e. the adjoint is the inverse of the forward method with a negative time
step. Using this definition, we can deduce that any explicit scheme will have an
implicit adjoint. This is the case for the results we have obtained with the
explicit Euler scheme in the previous section, albeit in a slightly convoluted
way. Specifically, if the original scheme is explicit, the adjoint scheme will
be implicit when integrated forward in time, but explicit when integrated
backwards in time (see~\cite{H2000} for how to reverse time in a Runge-Kutta
method).

To take advantage of different time integration methods, we first write the
semi-discrete variant of~\eqref{eq:burgers} as a coupled system of ODEs:
\[
\frac{\mathrm{d} \vect{u}}{\mathrm{d} t} = \mathcal{A}(\vect{u}),
\]
where $\mathcal{A}$ is the same as in~\eqref{eq:discrete_space_operator}. The
interdependence of the ODEs originates from the fact that $\mathcal{A}$ contains
a non-local discrete spatial derivative. We can write the system more explicitly,
for the case of the 3 cell stencil in the modified Lax-Friedrichs scheme, as:
\begin{equation} \label{eq:semidiscrete}
\frac{\mathrm{d} u_i}{\mathrm{d} t} = \mathcal{A}_i(u_{i - 1}, u_i, u_{i + 1}).
\end{equation}

First we will look at the second-order explicit Runge-Kutta method known as
Heun's method (or the explicit trapezoidal rule):
\begin{equation} \label{eq:rk_heun}
\left\{
\begin{aligned}
\vect{u}^{m + 1} =~& \vect{u}^m + \frac{\Delta t}{2}(\vect{k}^m_1 + \vect{k}^m_2),
& & \\
\vect{k}^m_1 =~ & \mathcal{A}(\vect{U}_1), & & \text{where }
    \vect{U}_1 = \vect{u}^m, \\
\vect{k}^m_2 =~ & \mathcal{A}(\vect{U}_2), & & \text{where }
    \vect{U}_2 = \vect{u}^m + \Delta t\, \vect{k}^m_1,
\end{aligned}
\right.
\end{equation}
which corresponds to the Butcher tableau with $b_1 = b_2 = \sfrac{1}{2}$,
$c_1 = 0, c_2 = 1, a_{11} = a_{12} = a_{22} = 0$ and $a_{21} = 1$. To find the
adjoint of this method, we will use the Lagrangian formalism where:
\[
\begin{aligned}
\mathcal{L}^h(\vect{u}, \vect{k}_1, \vect{k}_2,
              \vect{p}, \vect{l}_1, \vect{l}_2, \vect{g}) =
\mathcal{J}^h(\vect{u}, \vect{g})~- &
\sum_{m = 0}
<\vect{p}^{m + 1}, \left(
\vect{u}^{m + 1} - \vect{u}^m - \frac{\Delta t}{2}(\vect{k}^m_1 + \vect{k}^m_2)
\right)> \\
- &
\sum_{m = 0}
<\vect{l}^{m + 1}_1, (\vect{k}^m_1 - \mathcal{A}(\vect{U}_1))> \\
- &
\sum_{m = 0}
<\vect{l}^{m + 1}_2, (\vect{k}^m_2 - \mathcal{A}(\vect{U}_2))>,
\end{aligned}
\]
where the intermediate steps $\vect{k}_i$ are seen as independent variables
with constraints enforced by the new adjoint variables $\vect{l}_i$. To
find the adjoint Runge-Kutta scheme, we will differentiate the Lagrangian with
respect to the state variables $\vect{u}, \vect{k}_1$ and $\vect{k}_2$ to obtain
the following expressions for the gradients:
\[
\begin{aligned}
\frac{\partial \mathcal{L}^h}{\partial \vect{k}^m_1} =~ &
\frac{\Delta t}{2} \vect{p}^{m + 1} - \vect{l}^{m + 1}_1
+ \Delta t\left(\frac{\partial \mathcal{A}}{\partial \vect{u}^m}(\vect{U}_2)\right)^T \vect{l}^{m + 1}_2 = 0, \\
\frac{\partial \mathcal{L}^h}{\partial \vect{k}^m_2} =~ &
\frac{\Delta t}{2} \vect{p}^{m + 1} - \vect{l}^{m + 1}_2 = 0, \\
\frac{\partial \mathcal{L}^h}{\partial \vect{u}^m} =~ &
(-\vect{p}^{m} + \vect{p}^{m + 1}) +
\left(
\frac{\partial \mathcal{A}}{\partial \vect{u}^m}(\vect{U}_1)
\right)^T \vect{l}^{m + 1}_1 +
\left(
\frac{\partial \mathcal{A}}{\partial \vect{u}^m}(\vect{U}_2)
\right)^T \vect{l}^{m + 1}_2 = 0. \\
\end{aligned}
\]

The adjoint system we have obtained is another explicit Runge-Kutta method
integrated backwards in time with a Butcher tableau where the $a_{ij}$ matrix
of coefficients has been transposed. The resulting system can be written as:
\begin{equation} \label{eq:rk_heun_adjoint}
\left\{
\begin{aligned}
\vect{p}^m = & \vect{p}^{m + 1} +
\frac{\Delta t}{2} (\vect{l}^{m + 1}_1 + \vect{l}^{m + 1}_2), & & \\
\vect{l}^{m + 1}_1 = &
\left(
\frac{\partial \mathcal{A}}{\partial \vect{u}^m}(\vect{U}_1)
\right)^T \vect{P}_1, & & \text{where }
    \vect{P}_1 = \vect{p}^{m + 1} + \Delta t\, \vect{l}^{m + 1}_2, \\
\vect{l}^{m + 1}_2 = &
\left(
\frac{\partial \mathcal{A}}{\partial \vect{u}^m}(\vect{U}_2)
\right)^T \vect{P}_2, & & \text{where }
    \vect{P}_2 = \vect{p}^{m + 1},
\end{aligned}
\right.
\end{equation}
where we have redefined the original adjoint variables as follows:
\[
\vect{l}^{m + 1}_1 \equiv \frac{2}{\Delta t}
\left(
\frac{\partial \mathcal{A}}{\partial \vect{u}^m}(\vect{U}_1)
\right)^T \vect{l}^{m + 1}_1
\quad \text{and} \quad
\vect{l}^{m + 1}_2 \equiv \frac{2}{\Delta t}
\left(
\frac{\partial \mathcal{A}}{\partial \vect{u}^m}(\vect{U}_2)
\right)^T \vect{l}^{m + 1}_2.
\]

For general formulations of the adjoint of an arbitrary discrete $s$-stage
Runge-Kutta method see~\cite{S2006, H2000}. A very important property of the
adjoint, as defined in~\cite{H2000}, is that the order of the adjoint method
is the same as that of the original method, even when the original method is
explicit. The result we have obtained here agrees very well with the general
formulation from~\cite{H2000}.

For comparison with the adjoint method derived
previously~\eqref{eq:rk_heun_adjoint}, we will use an implicit method defined
using the methodology from~\cite{H2000}, namely the implicit midpoint method.
The implicit midpoint method is a 1-stage second-order symmetric Runge-Kutta
defined as:
\begin{equation} \label{eq:rk_midpoint}
\left\{
\begin{aligned}
\vect{u}^{m + 1} =~& \vect{u}^m + \Delta t \vect{k}^m_1,
& & \\
\vect{k}^m_1 =~ & \mathcal{A}(\vect{U}_1), & & \text{where }
    \vect{U}_1 = \vect{u}^m + \frac{\Delta t}{2} \vect{k}^m_1,
\end{aligned}
\right.
\end{equation}
and its adjoint is similarly given by:
\begin{equation} \label{eq:rk_midpoint_adjoint}
\left\{
\begin{aligned}
\vect{p}^m =~& \vect{p}^{m + 1} + \Delta t\, \vect{l}^m_1,
& & \\
\vect{l}^m_1 =~ &
\left(
\frac{\partial \mathcal{A}}{\partial \vect{u}^m}(\vect{U}_1)
\right)^T \vect{P}_1,
& & \text{where }
    \vect{P}_1 = \vect{p}^{m + 1} + \frac{\Delta t}{2} \vect{l}^m_1.
\end{aligned}
\right.
\end{equation}

\begin{table}[ht!]
\caption{Runge-Kutta: $L_1$ error $\log$ results for the adjoint variable $p$.}
\def\arraystretch{1.2}
\begin{center}
\begin{tabular}{lccccc}
  & 128 & 256 & 512 & \textbf{Adjoint Order} & \textbf{Stat Order} \\\hline\hline
Heun's Method
    & -1.04299 & -1.39669 & -1.74686 & \textbf{0.507} & \textbf{0.999} \\
Implicit Midpoint Method
    & -1.04384 & -1.39697 & -1.74701 & \textbf{0.507} & \textbf{0.999} \\\hline
\end{tabular}
\label{tbl:rk2_convergence}
\end{center}
\end{table}

We are mainly interested in comparing the two methods in the context of time
consistency. We have seen in the previous section (and
Figure~\ref{fig:ulbrich_time_wrong}) that an inconsistent time stepping method
may compromise the convergence of the numerical approximation to the correct
solution.

We will use the same modified Lax-Friedrichs scheme for the space discretization
and the exact solution given by~\eqref{eq:burgers_adjoint_exact_1} with a CFL
$= 0.9$ as before. We can see in Table~\ref{tbl:rk2_convergence} that the two
methods are basically indistinguishable both in error and final order. Note that
the order of the method remains $\approx 1$ for the state equations and
$\approx 0.5$ for the adjoint since the space discretization dominates at order
$\alpha$. This is not a major issue because we are simply interested in the
consistency and the convergence of the adjoint Runge-Kutta methods and not in
improving the accuracy.

\begin{figure}[ht!]
\centering
\begin{minipage}{0.5\textwidth}
\centering
\includegraphics[width=1.1\linewidth]{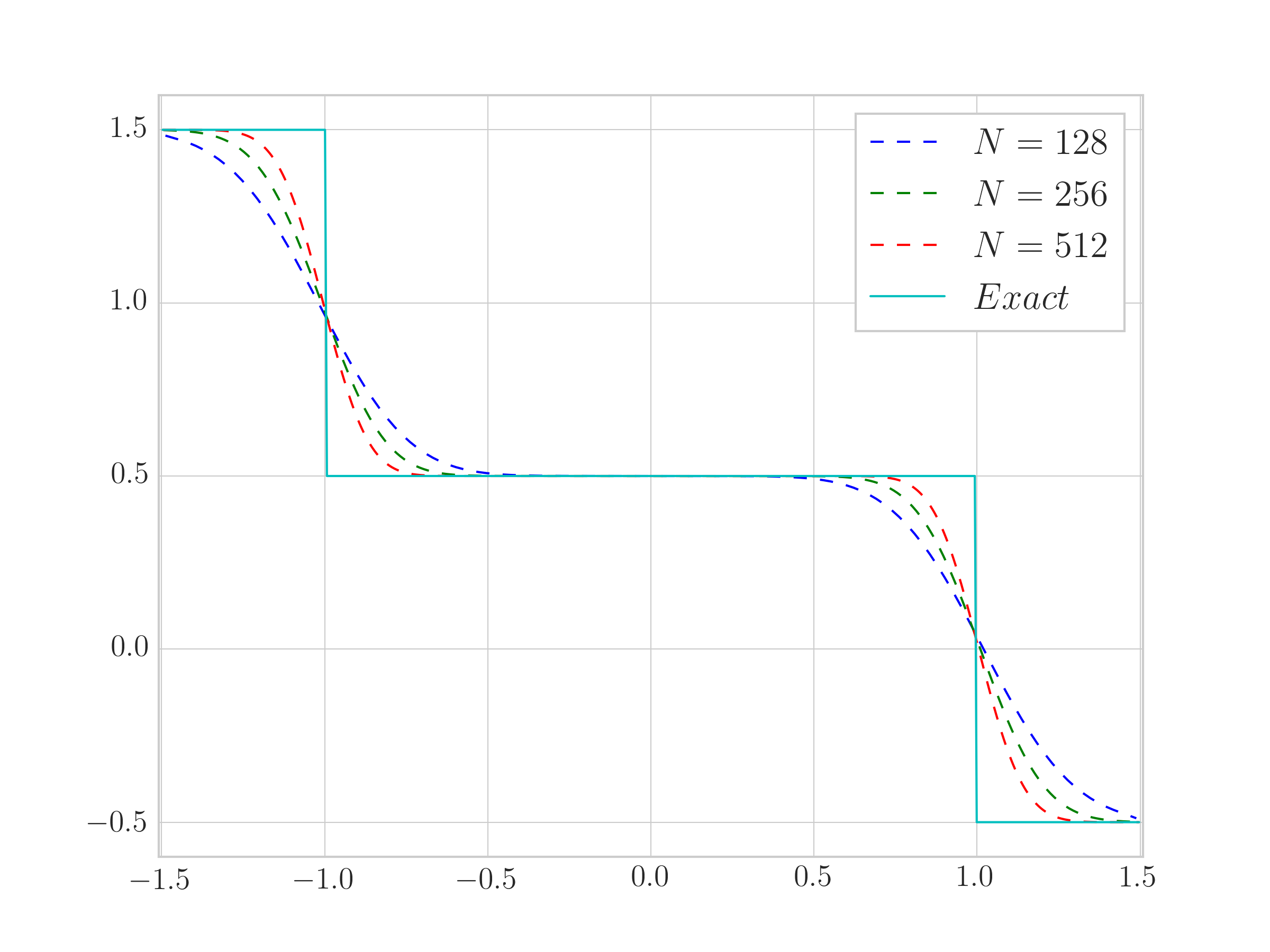}
\end{minipage}%
\begin{minipage}{0.5\textwidth}
\centering
\includegraphics[width=1.1\linewidth]{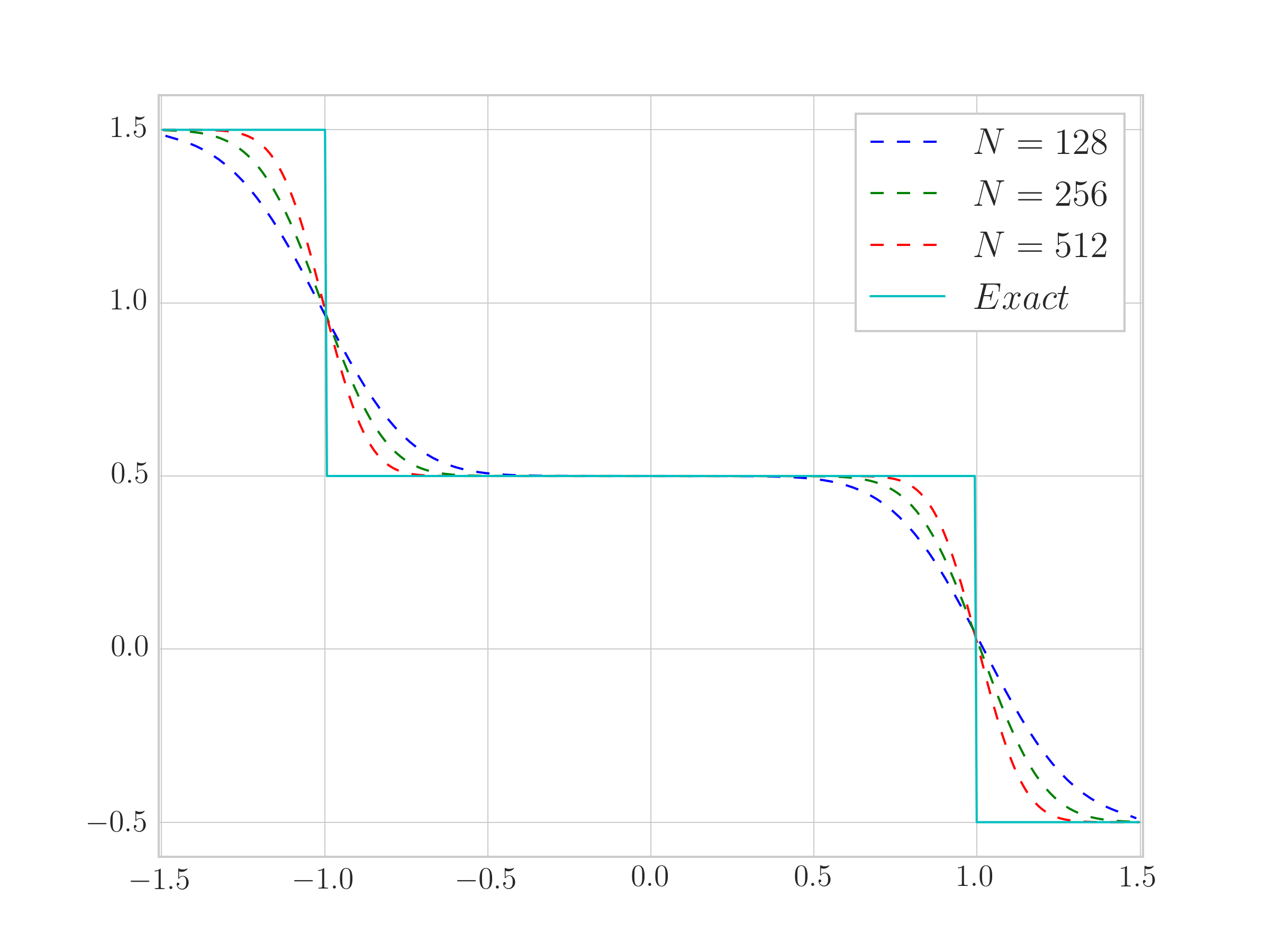}
\end{minipage}
\caption{Adjoint solution at $t = 0$ for~\eqref{eq:burgers_adjoint_exact_1}
using the explicit Heun's method (left) and the Implicit Midpoint method
(right) with consistent time-stepping.}
\label{fig:rk2_time_correct}
\end{figure}

We can see the same results in Figure~\ref{fig:rk2_time_correct} where
both methods still converge to the exact solution around the shock position
$x = 0$ at time $t = 0$. Similar convergence issues as in
Figure~\ref{fig:ulbrich_time_wrong} appear if the state variable is used at the
inconsistent time $\vect{u}^{m + 1}$ in the adjoint Runge-Kutta
methods~\eqref{eq:rk_heun_adjoint} and~\eqref{eq:rk_midpoint_adjoint}.

\section{Spatial Consistency}
\label{sec:space}

In this section we will investigate the spatial consistency of adjoint schemes
and issues that may arise from incomplete differentiation, non-differentiable
Riemann solvers, etc. For the purpose of this study we choose a high-order
nonlinear numerical scheme based on flux limiters. Note that flux limited
schemes (as well as any MUSCL scheme) are not proven to converge to the
correct adjoint solution. This is because one of the main ingredients in
convergence proofs regarding nonlinear conservation laws with shocks is
Oleinik's One-Sided Lipschitz Condition (OSLC):
\[
\frac{u(x, t) - u(y, t)}{x - y} \le \frac{1}{t}
\]
and its discrete equivalent. See, for example,~\cite{BJ1998, GJ2000, CPZ2008, U2001}
for general proofs that heavily involve the OSLC. The only limiter that is known
to satisfy the OSLC is the \texttt{maxmod} limiter~\cite{BO1988}, which,
unfortunately, does not remove spurious oscillations from the solution
(specifically, it allows undershoots), rendering it unacceptable in practice.
In~\cite{U2001}, an iterative method is proposed to ensure that a given slope,
initially limited with any commonly used limiter, is modified in such a way
that the scheme satisfies the OSLC. However, this makes the limiter itself overly
complicated (both implementation-wise and from a computation overhead point of
view), thus, not practical. Even though there are
no formal proofs,~\cite{NTT1994} suggests that TVD MUSCL schemes are bounded in
$Lip^+$, even though they are not monotonically decreasing, as required by
the OSLC, leading to convergence in certain cases.

Given the remarks from~\cite{NTT1994} and the fact that flux limited schemes
are among the most used nonlinear high-order schemes, we have adopted them for the
analysis of spatial consistency. Flux limited schemes are constructed as
a combination of a low-order monotone scheme and a high-order oscillatory scheme,
with the limiter choosing between the two such that no spurious oscillations
are introduced near discontinuities. More exactly, the fluxes
from~\eqref{eq:burgers_fv} are now given by:
\begin{equation} \label{eq:limited_flux}
f^m_{i - \frac{1}{2}}(u^m_i, u^m_{i - 1}, s_{i-\sfrac{1}{2}}, r_{i-\sfrac{1}{2}})
= f^{LO}_{i - \frac{1}{2}} + \phi\left(r_{i - \sfrac{1}{2}}\right)
\left(f^{HI}_{i - \frac{1}{2}} - f^{LO}_{i - \frac{1}{2}}\right),
\end{equation}
where $\phi$ is known as a \emph{flux limiter}. A common choice, adopted here,
is to use the first-order \emph{upwind} flux and the
second-order \emph{Lax-Wendroff} flux as our two fluxes:
\begin{equation} \label{eq:limited_lohi}
\left\{
\begin{aligned}
f^{LO}_{i - \frac{1}{2}} =~ & (1 - s_{i - \sfrac{1}{2}}) f(u^m_i) +
                              s_{i - \sfrac{1}{2}} f(u^m_{i - 1}), \\
f^{HI}_{i - \frac{1}{2}} =~ &
\frac{1}{2}
\left(
[f(u^m_i) + f(u^m_{i - 1})] +
\frac{\Delta t}{\Delta x} u^m_{i - \frac{1}{2}} [f(u^m_i) - f(u^m_{i - 1})]
\right),
\end{aligned}
\right.
\end{equation}
where $u^m_{i - \sfrac{1}{2}} = f'(u^{m}_{i - 1})$, if $u^m_i = u^m_{i - 1}$, or
the shock velocity otherwise, given by:
\[
u^m_{i - \frac{1}{2}} = \frac{u^m_i + u^m_{i - 1}}{2}
\]
and $s_{i - \sfrac{1}{2}}$ is an upwinding coefficient, given by the
continuous sigmoid:
\begin{equation} \label{eq:upwinding}
s_{i - \frac{1}{2}} =
\frac{1}{1 + \exp\left(-u^m_{i - \sfrac{1}{2}} / \delta\right)},
\quad
\delta \ll 1.
\end{equation}

We are now left with defining the flux limiter. For the purpose of this study,
we will consider a differentiable and non-differentiable flux limiter, namely
the \texttt{van Albada} limiter and the \texttt{minmod} limiter, given by:
\[
\phi_{\mathrm{albada}}(r) = \frac{r^2 + r}{r + 1}
\quad \text{and} \quad
\phi_{\mathrm{minmod}}(r) = \max(0, \min(1, r)),
\]
where the slope ratio is given by:
\begin{equation} \label{eq:slope_ratio}
r_{i - \frac{1}{2}} =
(1 - s_{i - \frac{1}{2}}) \frac{u^m_{i + 1} - u^m_i}{u^m_{i} - u^m_{i - 1}} +
s_{i - \frac{1}{2}} \frac{u^m_{i - 1} - u^m_{i - 2}}{u^m_{i} - u^m_{i - 1}}.
\end{equation}

This completely defines our scheme. We note that the van Albada limiter was
selected to ensure the differentiability of the numerical scheme itself, with
respect to $u^m_i$, so that it can be meaningfully linearized. Complete
differentiability requires a differentiable upwinding function (in this case, a
sigmoid) and a differentiable flux limiter (the van Leer limiter is also a good
choice). For testing purposes, we have also defined a discontinuous limiter so
that we may investigate its effect on the adjoint scheme. Using these
definitions, the adjoint scheme can be defined exactly like
in~\eqref{eq:adjoint_fv}, but with different adjoint ``fluxes''. We will
investigate two separate flux formulæ: one for which we differentiate the
flux limiter $\phi$ (referred to as \emph{complete differentiation}) and one
that assumes that the flux limiter is not a function of the state variables $u^m_i$
and therefore removing the differentiability property (referred to as
\emph{incomplete differentiation}). A previous study~\cite{AP2012} suggests that
the discrepancies are expected to remain small. However, the tests
in~\cite{AP2012} have been performed on the more complicated 2D Euler
equations and may hide some subtleties.

Generically, the main difference between the adjoint ``fluxes'' for the
completely and incompletely differentiated schemes is the fact that they have
a stencil of 5 cells and 3 cells, respectively. In the incompletely differentiated
case, the fluxes are:
\[
\left\{
\begin{aligned}
f^{*, m + 1}_{i, -} =~ &
\left[
\frac{\partial f^m_{i - \sfrac{1}{2}}}{\partial u^m_i}(u^m_i, u^m_{i - 1})
\right]
(p^{m + 1}_i - p^{m + 1}_{i - 1}), \\
f^{*, m + 1}_{i, +} =~ &
\left[
\frac{\partial f^m_{i + \sfrac{1}{2}}}{\partial u^m_i}(u^m_{i + 1}, u^m_i)
\right]
(p^{m + 1}_{i + 1} - p^{m + 1}_i),
\end{aligned}
\right.
\]
where the fluxes are only functions of $u^m_i$, without taking into account
the upwinding $s_{i - \sfrac{1}{2}}$ and the slope ration $r_{i - \sfrac{1}{2}}$,
and the completely differentiated fluxes are:
\[
\left\{
\begin{aligned}
f^{*, m + 1}_{i, -} =~ &
\left[
\frac{\partial f^m_{i - \sfrac{3}{2}}}{\partial u^m_i}
(u^m_{i - 1}, u^m_{i - 2}, s_{i - \sfrac{3}{2}}, r_{i - \sfrac{3}{2}})
\right]
(p^{m + 1}_{i - 1} - p^{m + 1}_{i - 2})~ + \\
&
~\left[
\frac{\partial f^m_{i - \sfrac{1}{2}}}{\partial u^m_i}
(u^m_i, u^m_{i - 1}, s_{i - \sfrac{1}{2}}, r_{i - \sfrac{1}{2}})
\right]~
(p^{m + 1}_i - p^{m + 1}_{i - 1}), \\
f^{*, m + 1}_{i, +} =~ &
~\left[
\frac{\partial f^m_{i + \sfrac{1}{2}}}{\partial u^m_i}
(u^m_{i + 1}, u^m_i, s_{i + \sfrac{1}{2}}, r_{i + \sfrac{1}{2}})
\right]~
(p^{m + 1}_{i + 1} - p^{m + 1}_i)~ + \\
&
\left[
\frac{\partial f^m_{i + \sfrac{3}{2}}}{\partial u^m_i}
(u^m_{i + 1}, u^m_{i + 2}, s_{i + \sfrac{3}{2}}, r_{i + \sfrac{3}{2}})
\right]
(p^{m + 1}_{i + 2} - p^{m + 1}_{i + 1}),
\end{aligned}
\right.
\]
where the fluxes, besides being functions of $u^m_i$, also depend on the
upwinding function $s_{i - \sfrac{1}{2}}(u^m_i, u^m_{i - 1})$~\eqref{eq:upwinding}
and the slope ratio $r_{i - \sfrac{1}{2}} (u^m_{i + 1}, u^m_i, u^m_{i - 1},
u^m_{i - 2})$~\eqref{eq:slope_ratio}, which can be differentiated using the
chain rule. Using the fluxes above, the adjoint scheme~\eqref{eq:adjoint_fv}
is completely defined in both cases.

\subsubsection*{Complete Upwind Scheme}

For a first test, we take the flux limiter $\phi \equiv 0$, which gives the
classic first-order upwind scheme, and investigate the effect of the parameter
$\delta$ in~\eqref{eq:upwinding}. The test case is the same simple scenario
described in~\eqref{eq:burgers_exact_1}-\eqref{eq:burgers_adjoint_exact_1} with
a CFL of $0.4$.

\begin{figure}[ht!]
\centering
\begin{subfigure}[b]{0.5\textwidth}
\includegraphics[width=\textwidth]{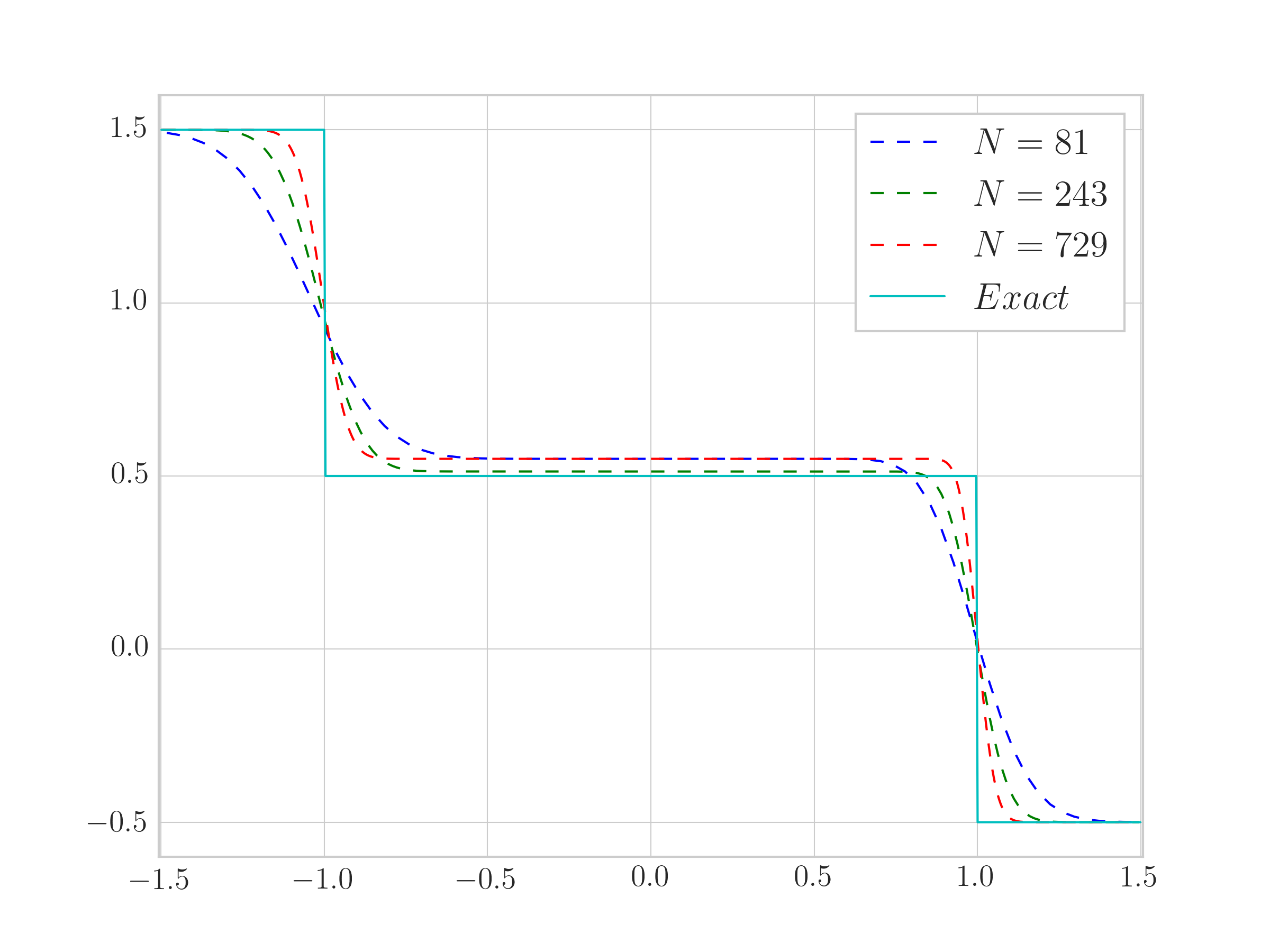}
\caption{} \label{fig:upwind_convergence}
\end{subfigure}%
\begin{subfigure}[b]{0.5\textwidth}
\includegraphics[width=\textwidth]{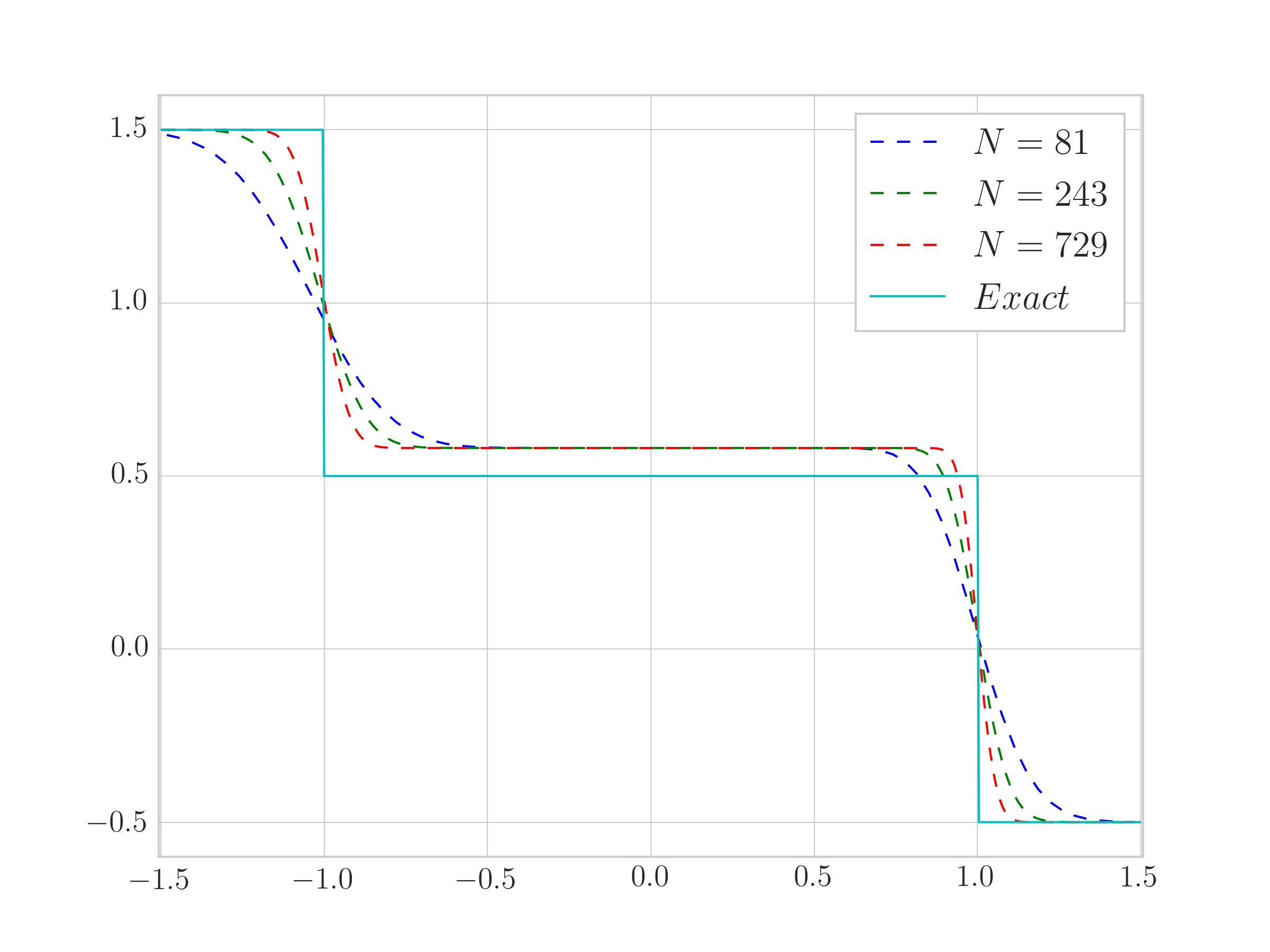}
\caption{} \label{fig:minmod_convergence}
\end{subfigure}

\begin{subfigure}[b]{0.5\textwidth}
\includegraphics[width=\textwidth]{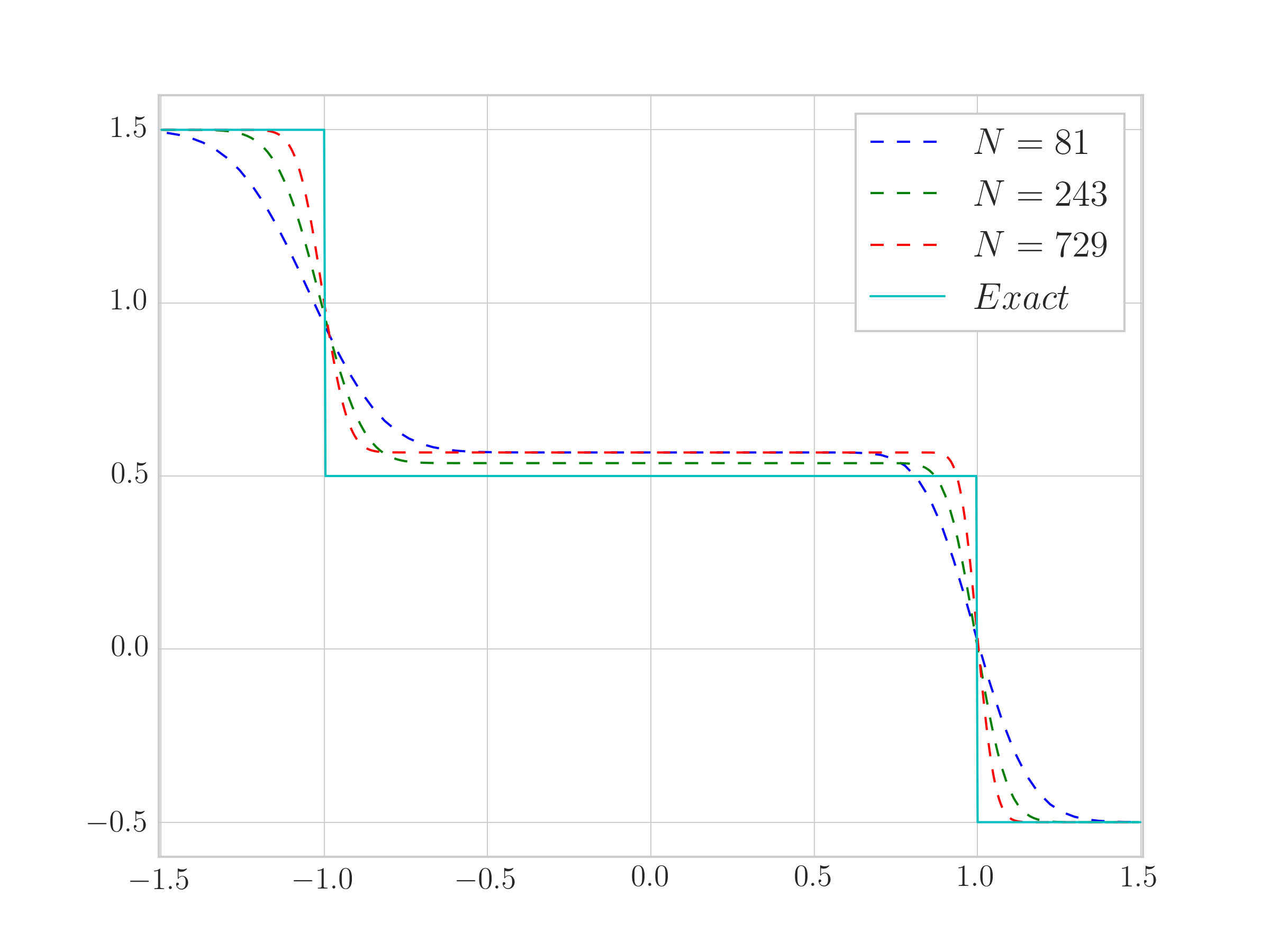}
\caption{} \label{fig:albada1_convergence}
\end{subfigure}%
\begin{subfigure}[b]{0.5\textwidth}
\includegraphics[width=\textwidth]{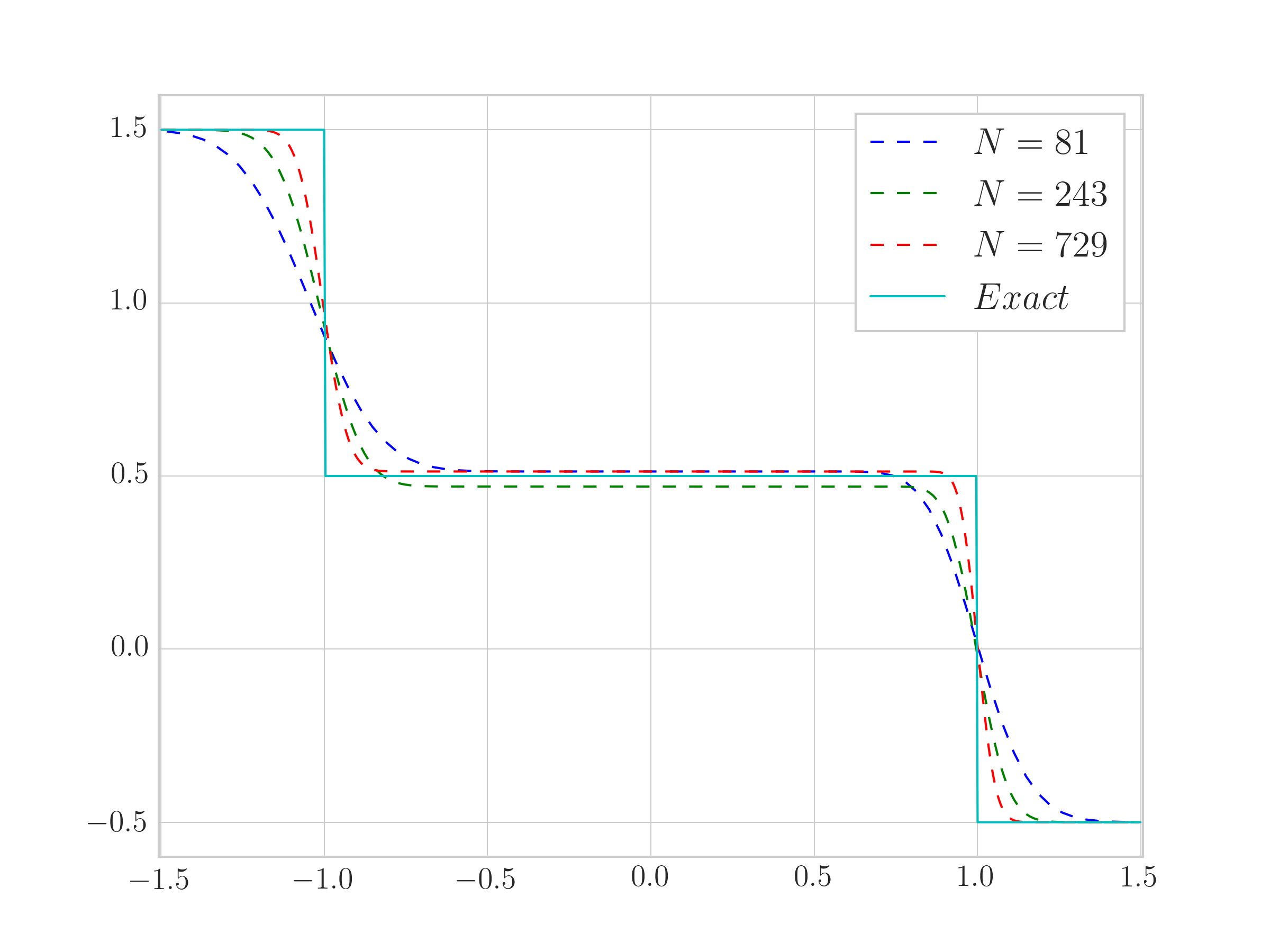}
\caption{} \label{fig:albada2_convergence}
\end{subfigure}
\caption{Convergence of the (a) upwind scheme, (b) a flux limited scheme with
the \texttt{minmod} limiter, (c) incompletely differentiated scheme with the
\texttt{van Albada} limiter and (d) completely differentiated scheme with the
\texttt{van Albada} limiter for the test case~\eqref{eq:burgers_adjoint_exact_1}
at time $t = 0$.}
\end{figure}

The results can be seen in Table~\ref{tbl:upwind_convergence} for various
values of $\delta$. For values of $\delta \ge 1$, the upwind schemes degenerates
to a centered (unstable) scheme, and the adjoint scheme becomes
unstable for CFL $0.4$. However, for values $< 0.1$ (even the discontinuous case of
$\delta = 0$) there are no significant differences in convergence order and
error. Results can also be seen in Figure~\ref{fig:upwind_convergence} for
$\delta = 0.001$ and are very similar in appearance to those obtained
in~\cite{AP2012}. Given these results, we expect the order to go to $0$ as the
grid is refined further since the scheme does not seem to converge.

\begin{table}[ht!]
\caption{Upwind Scheme: $L_1$ error $\log$ results for the adjoint
variable $p$.}
\def\arraystretch{1.2}
\begin{center}
\begin{tabular}{lccccc}
  & 81 & 243 & 729 &\textbf{Adjoint Order} & \textbf{State Order}\\\hline\hline
$\delta = 1.0$ \quad
    & -1.23692 & -1.61394 & -1.68173 & \textbf{0.202} & \textbf{1.000} \\
$\delta = 0.1$ \quad
    & -1.12534 & -1.80489 & -1.78244 & \textbf{0.299} & \textbf{0.999} \\
$\delta = 0.01$ \quad
    & -1.11337 & -1.83637 & -1.75990 & \textbf{0.294} & \textbf{0.999} \\
$\delta = 0.001$ \quad
    & -1.11223 & -1.80167 & -1.75728 & \textbf{0.293} & \textbf{0.999} \\
$\delta = 0.0$ \quad
    & -1.11223 & -1.80161 & -1.75728 & \textbf{0.293} & \textbf{0.999} \\\hline
\end{tabular}
\label{tbl:upwind_convergence}
\end{center}
\end{table}

\subsubsection*{Incomplete MINMOD Scheme}

Next, we consider the flux limited scheme with the \texttt{minmod} limiter.
By construction, the \texttt{minmod} limiter is not differentiable, so we cannot
fully differentiate the scheme, thus we must treat the flux limiter $\phi(r_i)$
as a constant.

We can see in Figure~\ref{fig:minmod_convergence} (CFL $= 0.3$) that the flux
limited scheme with the \texttt{minmod} limiter does not actually converge to
the exact solution, even though it does seem to converge to another solution
with very similar characteristics (i.e. continuous around the shock at $x = 0$).
Furthermore, the results from Table~\ref{tbl:minmod_convergence} show, for
various CFL conditions, the convergence order for the state and adjoint
equations. The state equations converge at first-order, as expected, but
the adjoint equations have an order that tends to $0$ as the grid is refined
(since it converges to the wrong solution).

\begin{table}[ht!]
\caption{\texttt{MINMOD}: $L_1$ error $\log$ results for the adjoint
variable $p$.}
\def\arraystretch{1.2}
\begin{center}
\begin{tabular}{lccccc}
  & 81 & 243 & 729 &\textbf{Adjoint Order} & \textbf{State Order}\\\hline\hline
CFL $= 0.6$ \quad
    & -1.44298 & -1.97570 & -2.45905 & \textbf{0.462} & \textbf{0.999} \\
CFL $= 0.3$ \quad
    & -0.89866 & -1.16698 & -1.35208 & \textbf{0.206} & \textbf{0.999} \\
CFL $= 0.15$ \quad
    & -0.81514 & -1.07005 & -1.24566 & \textbf{0.195} & \textbf{0.999} \\\hline
\end{tabular}
\label{tbl:minmod_convergence}
\end{center}
\end{table}

\subsubsection*{Complete and Incomplete van Albada Scheme}

Finally, we will look at the effects of complete vs. incomplete differentiation
of the flux functions in the differentiable flux limited scheme using the
\texttt{van Albada} limiter.

\begin{table}[ht!]
\caption{van Albada: $L_1$ error $\log$ results for the adjoint variable
$p$.}
\def\arraystretch{1.2}
\begin{center}
\begin{tabular}{lccccc}
  & 81 & 243 & 729 &\textbf{Adjoint Order} & \textbf{State Order}\\\hline\hline
Complete. CFL $= 0.4$ \quad
    & -1.32960 & -1.66087 & -2.24946 & \textbf{0.418} & \textbf{0.999} \\
Complete. CFL $= 0.2$ \quad
    & -1.14754 & -1.54951 & -1.89077 & \textbf{0.338} & \textbf{0.999} \\
Incomplete. CFL $= 0.4$ \quad
    & -1.02991 & -1.58054 & -1.57674 & \textbf{0.248} & \textbf{0.999} \\
Incomplete. CFL $= 0.2$ \quad
    & -0.89631 & -1.18067 & -1.38451 & \textbf{0.222} & \textbf{0.999} \\\hline
\end{tabular}
\label{tbl:albada_convergence}
\end{center}
\end{table}

The results of this test can be seen in Table~\ref{tbl:albada_convergence}. It
is clear that the incomplete differentiation of the \texttt{van Albada} limiter
has essentially the same problem as we have seen with the \texttt{minmod}
limiter and does not converge to the exact solution (see also
Figure~\ref{fig:albada1_convergence}). However, the completely differentiated
scheme does not seem to behave any better. Finally, neither exhibits
grid convergence properties at the intermediate state around the shock. In a
way, we can consider this as validation of the results from~\cite{AP2012}
which also mention that complete vs. incomplete differentiation does not have
a large impact on the end result.

Similar results can be seen in Table~\ref{tbl:albada_heun_convergence}, where we
have tested the van Albada limited scheme with the second-order Runge-Kutta
method described in~\eqref{eq:rk_heun}-\eqref{eq:rk_heun_adjoint}. The test,
using a CFL of $0.5$, has a slightly higher order for the tested grids, but
the solution is not grid converged and there is still no significant difference
between the completely and incompletely differentiated fluxes.

\begin{table}[ht!]
\caption{van Albada with Heun's time-stepping method: $L_1$ error $\log$ results
for the adjoint variable $p$.}
\def\arraystretch{1.2}
\begin{center}
\begin{tabular}{lccccc}
  & 81 & 243 & 729 &\textbf{Adjoint Order} & \textbf{State Order}\\\hline\hline
Complete \quad
    & -0.90772 & -1.18163 & -2.12105 & \textbf{0.552} & \textbf{0.869} \\
Incomplete \quad
    & -0.70418 & -0.92006 & -1.58714 & \textbf{0.401} & \textbf{0.869} \\
\end{tabular}
\label{tbl:albada_heun_convergence}
\end{center}
\end{table}

The same problem can be seen in all of the test cases we have seen in this
section. As mentioned in~\cite{G2002}, the adjoint numerical scheme basically
needs to approximate the boundary condition:
\[
\frac{\jump{G(u)}}{\jump{u}}
\]
at the shock and none of the schemes manage to capture it correctly for various
grid sizes and CFL conditions. This can probably be best seen in
Table~\ref{tbl:minmod_convergence} where for CFL $0.6$ the scheme converges
properly, but for CFL $0.3$ the boundary condition is no longer correctly
captured and the convergence suffers as an effect. We can, thus, conclude that
it is relatively easy to come across a grid and time step combination for which
the scheme converges to the wrong weak solution.

\begin{figure}[ht!]
\centering
\includegraphics[width=0.6\linewidth]{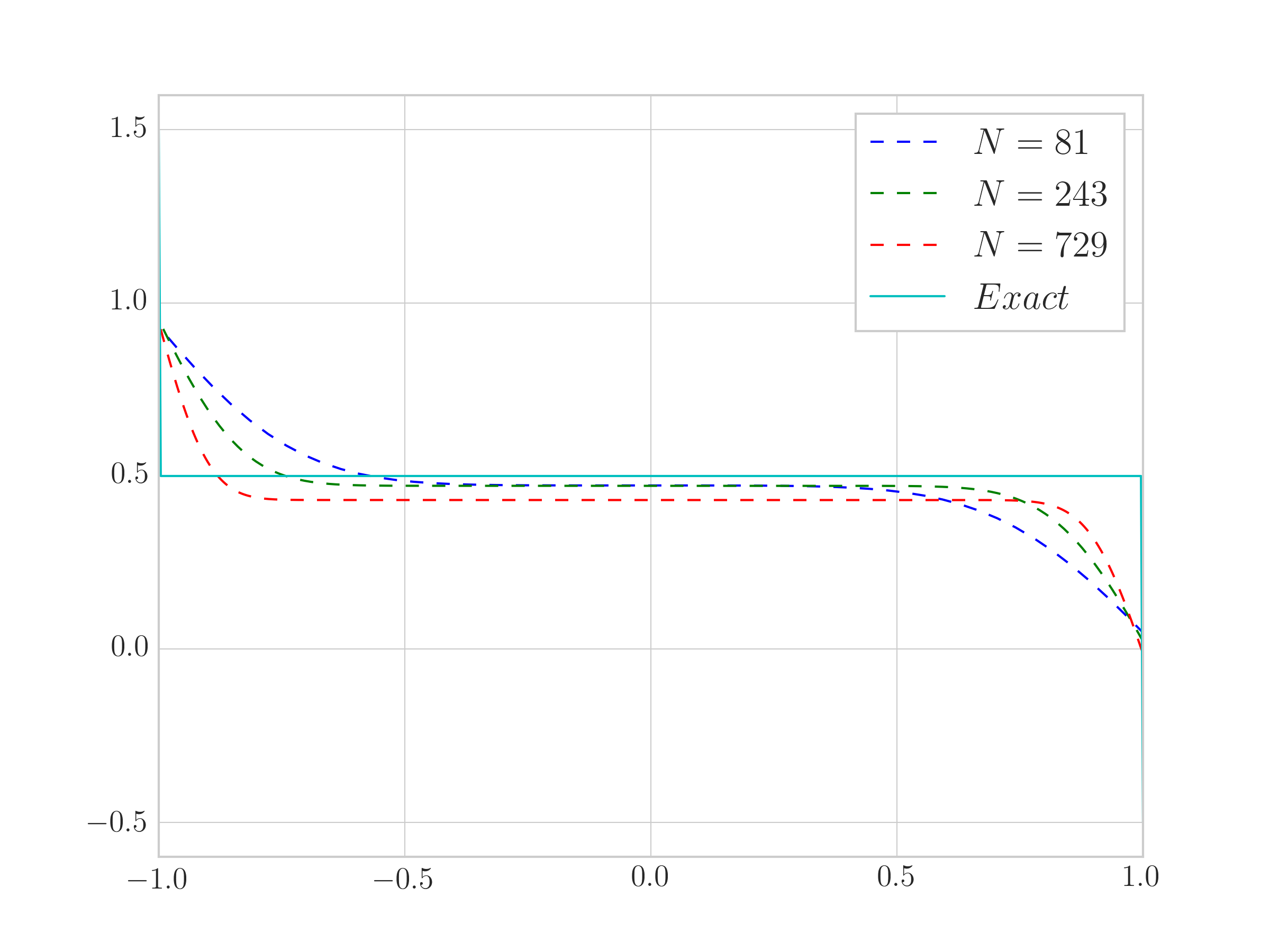}
\caption{Adjoint solution at $t = 0$ for~\eqref{eq:burgers_adjoint_exact_1}
using the modified Lax-Friedrichs scheme and the exact
solution~\eqref{eq:burgers_exact_1} for the state variable $\vect{u}^m$.}
\label{fig:ulbrich_exact_forward}
\end{figure}

This can even be seen with the modified Lax-Friedrichs scheme that normally
converges to the correct solution. For a simple test, we will use the known
exact solution~\eqref{eq:burgers_exact_1} as input for the adjoint equation
given by~\eqref{eq:adjoint_fv}. Using the exact solution means that the shock
is not smeared at all during the forward solve and the result can be clearly
seen in Figure~\ref{fig:ulbrich_exact_forward}. So, even for a converging
scheme as the modified Lax-Friedrichs scheme, if the shock is not sufficiently
smeared during the forward solve, the adjoint will not converge to the
correct solution for the same reasons as the flux limited schemes.

\section{Hybrid Numerical Scheme}
\label{sec:schemes}

In this section we will attempt to construct a higher order scheme for
Burgers' equation that also converges to the correct solution. Following
well-established methodologies~\cite{H1978}, the scheme we will look at is a
\emph{hybrid scheme} that uses a high-order flux limited scheme in smooth
regions and the modified Lax-Friedrichs scheme around the shock.
Using~\eqref{eq:limited_flux} and~\eqref{eq:lax_flux} we can construct a new
scheme as follows:
\begin{equation} \label{eq:mixed_flux}
f^m_{i - \frac{1}{2}} =
\begin{cases}
f^{LAX}_{i - \frac{1}{2}} = \frac{1}{2}(f(u^m_i) + f(u^m_{i - 1})) -
    \frac{\epsilon}{\Delta x} (u^m_i - u^m_{i - 1}),
    & \quad \rho_{i - \sfrac{1}{2}} > \rho, \\
f^{LIM}_{i - \frac{1}{2}} = f^{LO}_{i - \frac{1}{2}} + \phi\left(r_i\right)
    \left(f^{HI}_{i - \frac{1}{2}} - f^{LO}_{i - \frac{1}{2}}\right),
    & \quad \rho_{i - \sfrac{1}{2}} \le \rho,
\end{cases}
\end{equation}
where $\rho_{i - \sfrac{1}{2}}$ is a shock detector and $\rho > 0$ is a
given threshold. We have chosen to use a variant of Harten's
detector~\cite{H1978}:
\[
\rho_{i - \frac{1}{2}} = \max(\rho_i, \rho_{i - 1}),
\]
where:
\[
\rho_i =
\left|
\frac{
|u^m_{i + \sigma} - u^m_i| - |u^m_i - u^m_{i - \sigma}|}{
|u^m_{i + \sigma} - u^m_i| + |u^m_i - u^m_{i - \sigma}|}
\right|^r,
\]
where $\sigma$ is a shift parameter and $r$ is chosen such that the order
of the detector $\rho_{i - \sfrac{1}{2}} = \mathcal{O}(\Delta x^r)$ is
larger or equal to that of the high-order flux in the scheme (in this case,
$r \ge 1$). Note that, since the detector $\rho_{i - \sfrac{1}{2}} \in [0, 1]$,
Harten~\cite{H1978} originally defined the scheme as a convex combination
between the two fluxes, while we have chosen to use it as a hard switch.
Alternative detectors can be investigated, such as Jameson's detector~\cite{J1981},
a simple detector from~\cite{AP2012} that differentiates between shocks and
rarefactions or more recent detectors~\cite{OLLL2009} that can even distinguish
between discontinuities and highly oscillatory solutions.

The adjoint scheme is obtained by differentiating the fluxes as
in~\eqref{eq:adjoint_fv_general}. However, given that both the shock detector
and the scheme~\eqref{eq:mixed_flux} are discontinuous, we will not differentiate
them, so that the derivative is defined as follows:
\[
\frac{\partial f^m_{i - \sfrac{1}{2}}}{\partial u^m_i} =
\left\{
\begin{aligned}
\frac{\partial f^{LAX}_{i - \sfrac{1}{2}}}{\partial u^m_i},
    & \quad \rho_{i - \frac{1}{2}} > \rho, \\
\frac{\partial f^{LIM}_{i - \sfrac{1}{2}}}{\partial u^m_i}
    & \quad \rho_{i - \frac{1}{2}} \le \rho.
\end{aligned}
\right.
\]

We have seen in the previous section that complete vs. incomplete differentiation
makes little difference when the scheme does not converge, so we do not expect
the results to be affected. However, we still expect the
adjoint scheme to converge to the correct intermediate state around the shock
position because of the modified Lax-Friedrichs scheme. Using the exact solution
described in~\eqref{eq:burgers_exact_1}-\eqref{eq:burgers_adjoint_exact_1}
with a CFL $ = 0.5$, we will perform a number of tests on this new scheme to
see how it performs compared to the others:
\begin{description}
    \item[Test 1:] Complete differentiation of the \texttt{van Albada}
    limiter with $\rho = 0.3, \sigma = 2$ and $r = 1$.
    \item[Test 2:] Incomplete differentiation of the \texttt{van Albada}
    limiter with $\rho = 0.3, \sigma = 2$ and $r = 1$.
    \item[Test 3:] Complete differentiation of the \texttt{van Albada}
    limiter with $\rho = 0.3, \sigma = 2$ and $r = 2$.
    \item[Test 4:] Incomplete differentiation of the \texttt{van Albada}
    limiter with $\rho = 0.3, \sigma = 2$ and $r = 2$.
    \item[Test 5:] Incomplete differentiation of the \texttt{van Albada}
    limiter with $\rho = 0.1, \sigma = 1$ and $r = 1$.
    \item[Test 6:] Incomplete differentiation of the \texttt{van Albada}
    limiter with $\rho = 0.1, \sigma = 4$ and $r = 1$.
    \item[Test 7:] Incomplete differentiation of the \texttt{van Albada}
    limiter with $\rho = 0.5, \sigma = 2$ and $r = 1$.
    \item[Test 8:] Incomplete differentiation of the \texttt{van Albada}
    limiter with $\rho\! =\! 0.01, \sigma = 2$ and $r = 1$.
\end{description}

The results of the various tests can be see in Table~\ref{tbl:hybrid_convergence}.
We can see from Tests 1-4, that complete vs. incomplete differentiation has no
effect on the scheme. However, given that the flux limited scheme operates
mostly in constant regions, we can not draw conclusions from these tests.
Furthermore, we can see that increasing the stencil using $\sigma$ in Test 5-6
has a generally beneficial effect for convergence, while the overall order
stays the same. Similar results can be seen when using a smaller threshold $\rho$,
i.e. using the high-order flux less, in Test 7-8.

Overall, the convergence results in Table~\ref{tbl:hybrid_convergence} are
slightly better than those of the previously tested schemes, suggesting that
better convergence for the adjoint equation is possible.

\begin{table}[ht!]
\caption{Hybrid Scheme: $L_1$ error results for the adjoint variables $e_p$.}
\def\arraystretch{1.2}
\begin{center}
\begin{tabular}{lccccc}
& 81 & 243 & 729 &\textbf{Adjoint Order} & \textbf{State Order}\\\hline\hline
Test 1 \quad
    & -0.94617 & -1.66985 & -2.33860 & \textbf{0.633} & \textbf{0.993} \\
Test 2 \quad
    & -0.94792 & -1.68200 & -2.30667 & \textbf{0.618} & \textbf{0.993} \\\hline
Test 3 \quad
    & -0.86818 & -1.38219 & -1.81292 & \textbf{0.429} & \textbf{1.003} \\
Test 4 \quad
    & -0.92688 & -1.54892 & -2.08032 & \textbf{0.524} & \textbf{1.003} \\\hline
Test 5 \quad
    & -0.95526 & -1.69365 & -2.35130 & \textbf{0.635} & \textbf{1.002} \\
Test 6 \quad
    & -0.93081 & -1.67549 & -2.34335 & \textbf{0.642} & \textbf{1.002} \\\hline
Test 7 \quad
    & -0.92227 & -1.64159 & -2.22935 & \textbf{0.594} & \textbf{1.005} \\
Test 8 \quad
    & -0.94667 & -1.68759 & -2.34862 & \textbf{0.638} & \textbf{1.002} \\\hline
\end{tabular}
\label{tbl:hybrid_convergence}
\end{center}
\end{table}

To more thoroughly test the new scheme, we propose two new test cases, inspired
by~\cite{G2002}. We use the same cost function as in~\eqref{eq:cost} with
$G(u) = u^5 - u$. First, we take a stationary shock problem:
\begin{equation} \label{eq:burgers_exact_2}
u(x, t) =
\left\{
\begin{aligned}
& 1, & &
    \quad x < \min(0.25 + t, 0.5), \\
& \frac{x - 0.5}{t - 0.25}, & &
    \quad \min(0.25 + t, 0.5) < x < \max(0.75 - t, 0.5), \\
& -1, & &
    \quad x > \max(0.75 - t, 0.5),
\end{aligned}
\right.
\end{equation}
and the adjoint solution:
\begin{equation} \label{eq:burgers_adjoint_exact_2}
p(x, t) =
\begin{cases}
4, & \quad x < t, \\
0, & \quad t < x < 1 - t, \\
4, & \quad x > 1 - t.
\end{cases}
\end{equation}

The second test case is a moving shock problem with the exact solution:
\begin{equation} \label{eq:burgers_exact_3}
u(x, t) =
\left\{
\begin{aligned}
& 1, & &
    x < \min(0.25 + t, 0.475 + 0.1 t), \\
& \frac{x - 0.5}{t - 0.25}, & &
    \min(0.25 + t, 0.475 + 0.1 t) < x < \max(0.7 - 0.8t, 0.475 + 0.1t), \\
& -0.8, & &
    x > \max(0.7 - 0.8 t, 0.475 + 0.1 t),
\end{aligned}
\right.
\end{equation}
with the adjoint solution:
\begin{equation} \label{eq:burgers_adjoint_exact_3}
p(x, t) =
\begin{cases}
4,          & \quad x < 0.025 + t, \\
-0.2624,    & \quad 0.025 + t < x < 0.925 - 0.8 t, \\
1.048,      & \quad x > 0.925 - 0.8 t.
\end{cases}
\end{equation}

\begin{figure}[ht!]
\centering
\begin{subfigure}[b]{0.5\textwidth}
\includegraphics[width=\textwidth]{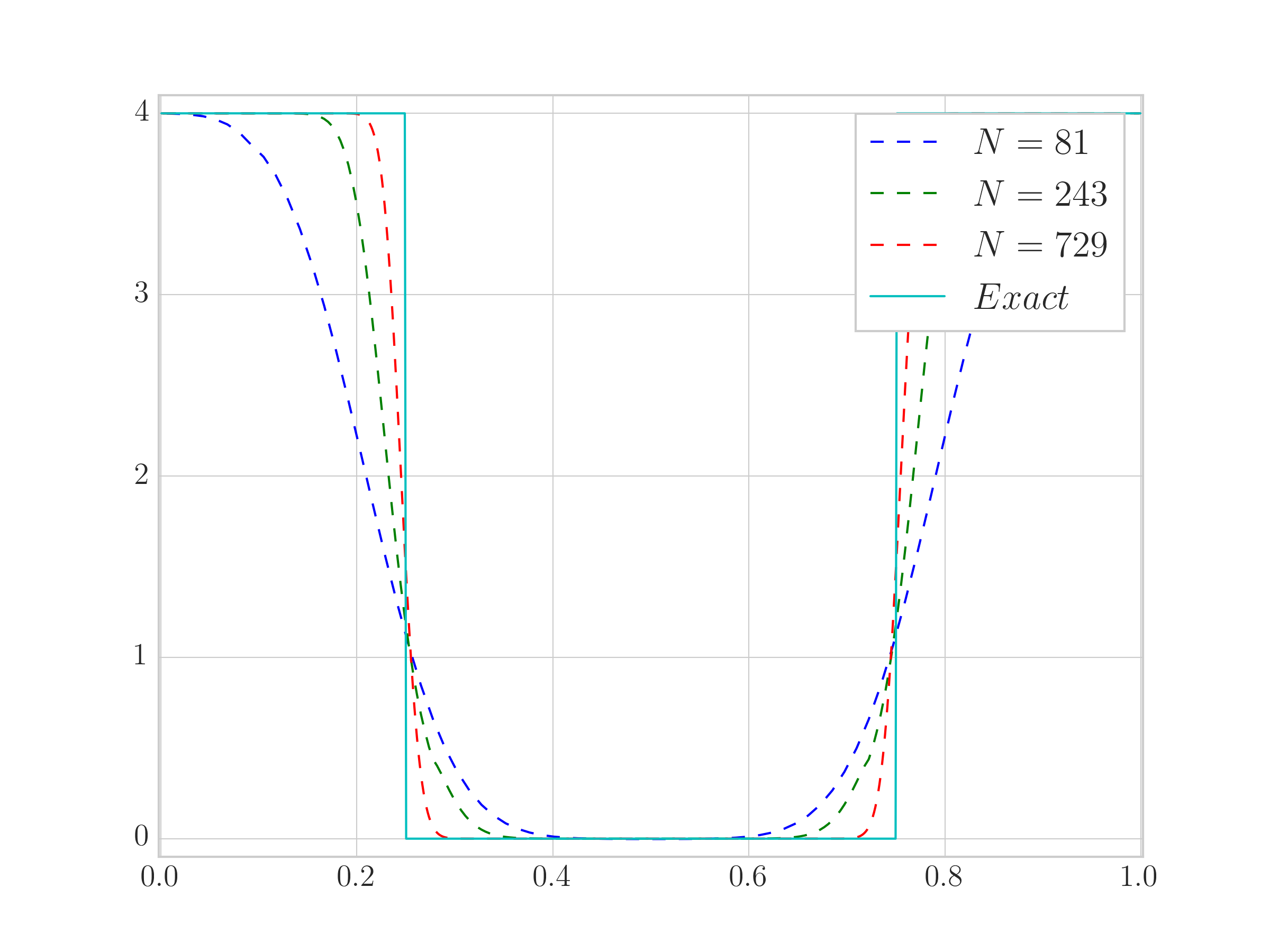}
\caption{} \label{fig:hybrid_sol2_convergence}
\end{subfigure}%
\begin{subfigure}[b]{0.5\textwidth}
\includegraphics[width=\textwidth]{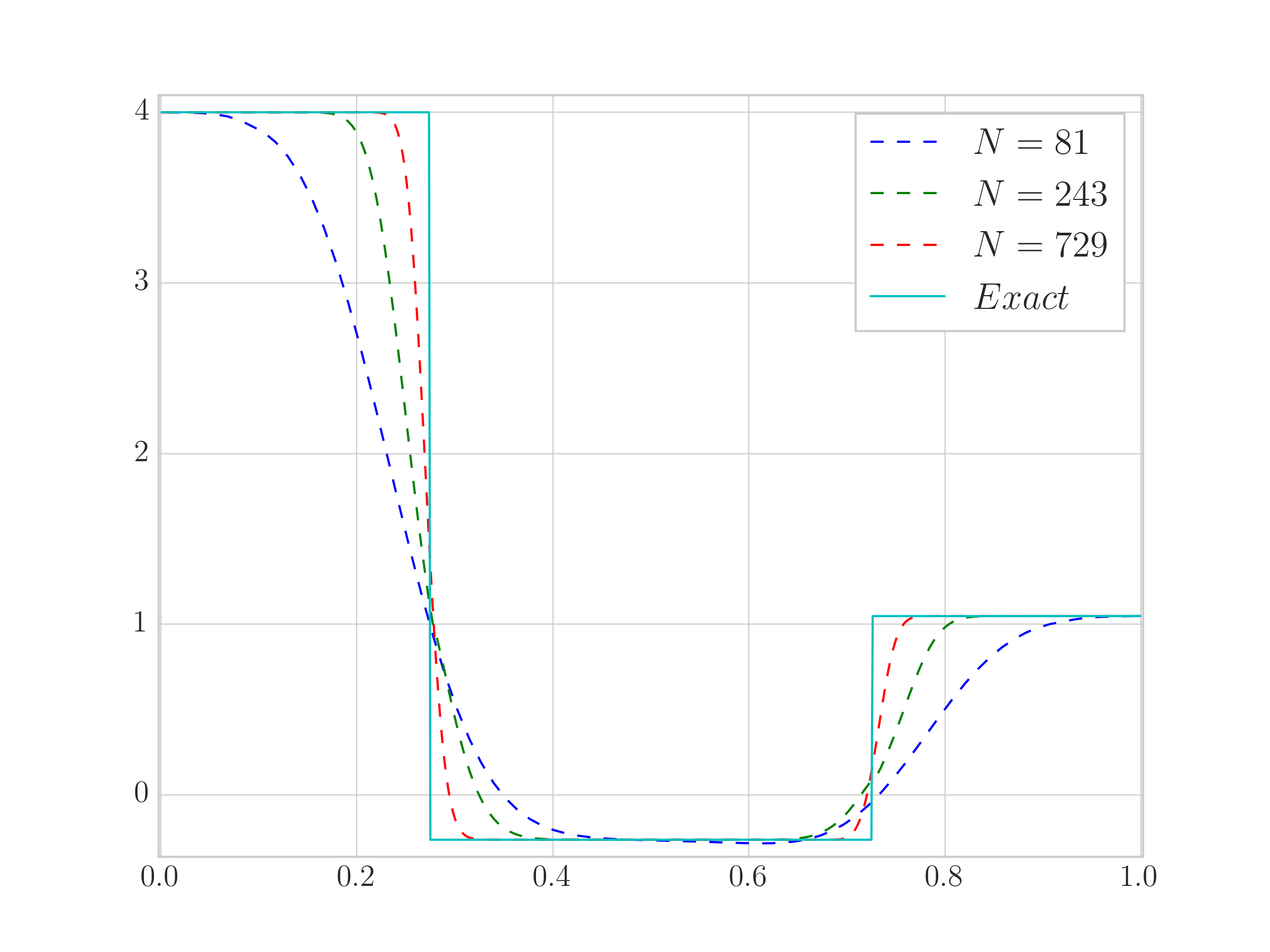}
\caption{} \label{fig:hybrid_sol3_convergence}
\end{subfigure}
\caption{(a) Approximate solutions to~\eqref{eq:burgers_adjoint_exact_2}
and (b) approximation solutions to~\eqref{eq:burgers_adjoint_exact_3}
at $T = 0.25$.}
\label{fig:hybrid_convergence}
\end{figure}

Test cases~\eqref{eq:burgers_exact_2}-\eqref{eq:burgers_adjoint_exact_2}
and~\eqref{eq:burgers_exact_3}-\eqref{eq:burgers_adjoint_exact_3} have
$T = 0.5$ and $\Omega = [0, 1]$. We can see the results in
Figure~\ref{fig:hybrid_convergence} where the scheme is shown to converge in
both cases.

The new hybrid scheme we have proposed seems to have marginally improved
accuracy, while maintaining the same convergence qualities as the original
modified Lax-Friedrichs scheme. However, it is very sensitive to the choice
of parameters $(\rho, \sigma)$, that define the width around the
shock where the Lax-Friedrichs scheme is used. We can see in the tests performed
earlier that $\rho$ has to be quite small for convergence, generally $< 0.1$
for the specific case of~\eqref{eq:burgers_adjoint_exact_1}. For $\gtrsim 0.5$
the scheme no longer converges, although, in our tests, it does not deviate
from the correct solution as much as the flux limited
scheme~\eqref{eq:limited_flux} alone.

\section{Conclusions}

In this paper we have analyzed numerical schemes for
scalar conservation laws in the presence of shocks. We have seen that a
grid converged numerical scheme, like the modified Lax-Friedrichs scheme, needs to
smear the shock across multiple cells as the domain is refined to correctly
capture the state around the shock position in the adjoint solution. This was not
the case for the flux limited schemes in Section~\ref{sec:space} and, as a
result, none of them converged to the correct solution for all space and time
discretizations. However, even in the case of non-converging numerical schemes,
in most cases, the approximate solution was not far from the desired exact
solution. We expect the error that occurs from incorrectly capturing the
intermediate state to depend heavily on the magnitude of the shock. This
implies that, for small shocks, a high-order, non-converging scheme may not
incur a sizable error and may still provide sufficient resolution. While
failure to converge may be a problem in sensitivity analysis, an iterative
optimization algorithm may afford a slower convergence rate at the cost of
higher accuracy in the numerical results.

While the error in convergence may prove manageable in certain scenarios, another
error has been considered in this work, namely the error that comes with
incompletely differentiating a highly nonlinear numerical scheme. We have
repeatedly seen, for various limiters and various time stepping schemes, that
incomplete differentiation also does not have much of an impact on the
result. Specifically, the convergence properties do not improve when completely
differentiating a nonlinear flux.

Finally, we have seen that a consistent time integration also has a significant
impact on the convergence of the numerical scheme. Namely, if we were to
discretize an adjoint obtained from the continuous state equations, it would
be unclear at what time $t^* \in [t^m, t^{m + 1}]$ the state variables should
be taken and what the impact is on the resulting scheme. This problem becomes
completely clear in the discrete case where we are left with a single choice
if we desire to be consistent with the discrete state equations. Higher order
numerical integrators have also been investigated, both implicit and explicit,
yielding favorable results. However, in our case, where the smooth regions are
constant, higher order time integration did not have any impact on the resulting
accuracy of the scheme.



\begin{thebibliography}{9}
\bibitem{P1974}
    O. Pironneau,
    \emph{On Optimum Design in Fluid Mechanics},
    Journal of Fluid Mechanics, Vol. 64, pp. 97-110, 1974.
\bibitem{J1988}
    A. Jameson,
    \emph{Aerodynamic Design via Control Theory},
    Journal of Scientific Computing, Vol. 3, 1988.
\bibitem{G2003}
    M. D. Gunzburger,
    \emph{Perspectives in Flow Control and Optimization},
    SIAM, 2003.
\bibitem{BS2012}
    A. Borzi, V. Schulz,
    \emph{Computational Optimization of Systems Governed by Partial
    Differential Equations},
    SIAM, 2012.
\bibitem{HPUU2008}
    M. Hinze, R. Pinnau, M. Ulbrich, S. Ulbrich,
    \emph{Optimization with PDE Constraints},
    Springer, 2008.
\bibitem{L1971}
    J. L. Lions,
    \emph{Optimal Control of Systems Governed by Partial Differential Equations},
    Springer-Verlag, 1971.
\bibitem{ZZANS2000}
    S. Zhang, X. Zhou, J. Ahlquist, I. M. Navon, J. G. Sela,
    \emph{Use of Differentiable and Non-differentiable Optimization Algorithms
    for Variational Data Assimilation with Discontinuous Cost Functions},
    Monthly Weather Review, Vol. 128, pp. 4031-4044, 2000.
\bibitem{IS1999}
    A. Iollo, M. D. Salas,
    \emph{Optimum Transonic Airfoils Based on Euler Equations},
    Computers and Fluids, Vol. 28, pp. 653-674, 1999.
\bibitem{HL2001}
    L. Huyse, R. M. Lewis,
    \emph{Aerodynamic Shape Optimization of Two-dimensional Airfoils Under
    Uncertain Conditions},
    Institute for Computer Applications in Science and Engineering Report, 2001.
\bibitem{GIMM1981}
    J. Glimm, E. Isaacson, D. Marchesin, O. McBryan,
    \emph{Front Tracking for Hyperbolic Systems},
    Advances in Applied Mathematics, Vol. 2, pp. 91-119, 1981.
\bibitem{MH1997}
    T. Matsuzawa, M. Hafez,
    \emph{Optimum Shape Design Using Adjoint Equations for Compressible Flows
    with Shock Waves},
    AIAA Paper, 97-2078, 1997.
\bibitem{OONN1997}
    A. Oyama, S. Obayashi, K. Nakahashi, T. Nakamura,
    \emph{Transonic Wing Optimization Using Genetic Algorithms},
    AIAA Paper, 97-1854, 1997
\bibitem{CHS1997}
    E. M. Cliff, M. Heinkenschloss, A. R. Shenov,
    \emph{An Optimal Control Problem for Flows with Discontinuities},
    Journal of Optimization Theory and Applications, Vol. 94, pp. 273-309, 1997.
\bibitem{BM1995}
    A. Bressan, A. Marson,
    \emph{A Variational Calculus for Discontinuous Solutions of Systems of
    Conservation Laws},
    Communications in Partial Differential Equations, Vol. 20, 1995.
\bibitem{HN2003}
    C. Homescu, I. M. Navon,
    \emph{Optimal Control of Flow with Discontinuities},
    Journal of Computational Physics, Vol. 187, pp. 660-682, 2003.
\bibitem{BP2003}
    C. Bardos, O. Pironneau,
    \emph{Derivatives and Control in the Presence of Shocks},
    Computational Fluid Dynamics Journal, pp. 383-392, 2003.
\bibitem{U2001}
    S. Ulbrich,
    \emph{Optimal Control of Nonlinear Hyperbolic Conservation Laws with
    Source Terms},
    Habilitation Thesis, Zentrum Mathematik, Technische Universitat Munchen,
    Germany, 2001
\bibitem{GU12010}
    M. B. Giles, S. Ulbrich,
    \emph{Convergence of Linearized and Adjoint Approximations for
    Discontinuous Solutions of Conservation Laws. Part 1: Linearized
    Approximations and Linearized Output Functionals},
    SIAM Journal of Numerical Analysis, Vol. 48, pp. 882-904, 2010.
\bibitem{GU22010}
    M. B. Giles, S. Ulbrich,
    \emph{Convergence of Linearized and Adjoint Approximations for
    Discontinuous Solutions of Conservation Laws. Part 2: Adjoint
    Approximations and Extensions},
    SIAM Journal of Numerical Analysis, Vol. 48, pp. 905-921, 2010.
\bibitem{G2002}
    M. B. Giles,
    \emph{Discrete Adjoing Approximations with Shocks},
    Proceedings of the Ninth International Conference on Hyperbolic Problems,
    pp.185-194, 2002.
\bibitem{GP2000}
    M. B. Giles, N. A. Pierce,
    \emph{An Introduction to the Adjoint Approach to Design},
    Flow, Turbulence and Combustion, Vol. 65, pp. 393-415, 2000.
\bibitem{AP2012}
    F. Alauzet, O. Pironneau,
    \emph{Continuous and Discrete Adjoints to the Euler Equations for Fluids},
    International Journal for Numerical Methods in Fluids, Vol. 70, pp. 135-157, 2012.
\bibitem{CPZ2008}
    C. Castro, F. Palacios, E. Zuazua,
    \emph{An Alternating Descent Method for the Optimal Control of the Inviscid
    Burgers Equation in the Presence of Shocks},
    Mathematical Models and Methods in Applied Science, Vol. 18, 2008.
\bibitem{FSS2012}
    M. Fosas de Pado, D. Sipp, P. J. Schimd,
    \emph{Efficient Evaluation of the Direct and Adjoint Linearized Dynamics
    from Compressible Flow Solvers},
    Journal of Computational Physics, Vol. 231, pp. 7739-755, 2012.
\bibitem{S2006}
    A. Sandu,
    \emph{On the Properties of Runge-Kutta Discrete Adjoints},
    Proceedings of the 6th international conference on Computational Science,
    Vol. IV, pp. 550-557, 2006.
\bibitem{H2000}
    W. W. Hager,
    \emph{Runge-Kutta Methods in Optimal Control and the Transformed Adjoint
    System},
    Numerische Mathematik, Vol. 87, pp. 247-282, 2000.
\bibitem{HWL2006}
    E. Hairer, G. Wanner, C. Lubich,
    \emph{Geometric Numerical Integration},
    Springer, 2006.
\bibitem{BO1988}
    Y. Brenier, S. Osher,
    \emph{The Discrete One-Sided Lipschitz Condition for Convex Scalar
    Conservation Laws},
    SIAM Journal of Numerical Analysis, Vol. 25, pp. 8-23, 1988.
\bibitem{NTT1994}
    H. Nessyahu, E. Tadmor, T. Tassa,
    \emph{The Convergence Rate of Godunov Type Schemes},
    SIAM Journal of Numerical Analysis, Vol. 31, pp. 1-16, 1994.
\bibitem{H1978}
    A. Harten,
    \emph{The Artificial Compression Method for Computation of Shocks and
    Contact Discontinuities: III. Self-Adjusting Hybrid Schemes},
    Mathematics of Computation, Vol. 32, pp. 363-289, 1978.
\bibitem{J1981}
    A. Jameson, W. Schmidt, E. Turkel,
    \emph{Numerical Solutions of the Euler Equations by Finite Volume Methods
    Using Runge-Kutta Time-Stepping Schemes},
    AIAA Paper 1259, 1981.
\bibitem{OLLL2009}
    M. Oliveria, P. Lu, X. Liu, C. Liu,
    \emph{Universal High Order Subroutine with New Shock Detector for Shock
    Boundary Layer Interaction},
    AIAA Paper 2009-1139, 2009.
\bibitem{NJ2000}
    S. Nadarajah, A. Jameson,
    \emph{A Comparison of the Continuous and Discrete Adjoint Approach to
    Automatic Aerodynamic Optimization},
    AIAA Paper 2000-0667, 38th Aerospace Sciences Meeting and Exhibit, 2000.
\bibitem{BJ1998}
    F. Bouchut, F. James,
    \emph{One-Dimensional Transport Equations with Discontinuous Coefficients}
    Nonlinear Analysis, Vol. 32, pp. 891-993, 1998.
\bibitem{GJ2000}
    L. Gosse, F. James,
    \emph{Numerical Approximations of One-Dimensional Linear Conservation
    Equations with Discontinuous Coefficients},
    Mathematics of Computation, Vol. 69, pp. 987-1015, 2000.
\end{thebibliography}
\end{document}